\documentstyle[12pt,epsf,axodraw]{article}
 \makeatletter
 \def\appendix{\par\clearpage
   \setcounter{section}{0}
   \setcounter{subsection}{0}
   \@addtoreset{equation}{section}

   \def\@sectname{Appendix~}
   \def\theequation{\thesection.\arabic{equation}}
   \def\thesection{\Alph{section}}}
 \makeatother

 \renewcommand{\theequation}{\thesection.\arabic{equation}}

\begin{document}
 \begin{titlepage}
 \hskip 11cm \vbox{ \hbox{BUDKERINP/2002-60}}

 \vskip 0.3cm \centerline{\bf ON THE CALCULATION OF THE NLO VIRTUAL
 PHOTON IMPACT FACTOR$^{~\ast}$}

 \vskip 0.8cm \centerline{  V.S. Fadin$^{a, b~\dagger}$, D.Yu.
 Ivanov$^{c, e~\ddagger}$ and M.I. Kotsky$^{a~\dagger\dagger}$}
 \vskip .3cm \centerline{\sl $^a$ Budker Institute for Nuclear
 Physics, 630090 Novosibirsk, Russia} \centerline{\sl $^b$
 Novosibirsk State University, 630090 Novosibirsk, Russia}
 \centerline{\sl $^c$ Institute of Mathematics, 630090
 Novosibirsk, Russia}
 \centerline{\sl $^e$ Regensburg University, Germany}
 \vskip 1cm

 \begin{abstract}
 The definition of the virtual photon impact factor involves the integration
 of the $s$-channel discontinuity of the photon-Reggeon scattering amplitude
 over the right cut. It permits to formulate a new approach for the calculation
 of the impact factor based on analytical properties of the amplitude in question.
 In the next-to-leading order it may give a possibility for considerable
 simplification of the calculation. We have shown that a part of the diagrams
 contributing to the impact factor can be treated without their
 real calculation.
 \end{abstract}
 \vfill
 \hrule
 \vskip.3cm
 \noindent
 $^{\ast}${\it Work supported in part by INTAS and in part by the
 Russian Fund of Basic Researches.}
 \vfill
 $ \begin{array}{ll}
 ^{\dagger}\mbox{{\it e-mail address:}} &
  \mbox{FADIN@INP.NSK.SU}\\
 \end{array}
 $

 $ \begin{array}{ll}
 ^{\ddagger}\mbox{{\it e-mail address:}} &
   \mbox{D-IVANOV@MATH.NSC.RU}
 \end{array}
 $

 $ \begin{array}{ll} ^{\dagger\dagger}\mbox{{\it e-mail address:}}
 &
  \mbox{KOTSKY@INP.NSK.SU}\\
 \end{array}
 $

 \vfill
 \vskip .1cm
 \vfill
 \end{titlepage}
 \eject

 \section{Introduction}
 \setcounter{equation}{0}

 There are at least two reasons why the total cross section of interaction of
 photons with large virtualities $\sim Q^2\gg \Lambda^2_{QCD}$  at high  c.m.s.
 energies $\sqrt s$ is an attractive object for theoretical investigation. The
 first one is that it was  experimentally measured (see \cite{OPAL} - \cite{ALEPH}
 and references therein); the second at large  enough  $Q^2$ this cross section
 can be calculated in a framework of perturbative QCD. The most interesting is
 the region of so small values of $x=Q^2/s$ where, firstly, the main contribution
 to the cross section is given by nondecreasing with energy terms, and secondly,
 a resummation of the higher order contributions
 enhanced by powers of
 $\log(1/x)$ is necessary.

 The most common basis for such resummation is given by the BFKL approach~\cite{FKL}.
 This approach in the leading logarithmic approximation (LLA),
 when only leading terms $(\alpha_s\ln s)^n$ are resumed,
 was extensively used for theoretical analysis of $\gamma^*\gamma^*$ interaction \cite{Bartels}.
 Unfortunately, in the LLA  neither scale of energy, nor argument of the running coupling
 constant $\alpha _{s}$ are fixed, so that for accurate
 theoretical prediction  we have to know the radiative corrections to the
 LLA. Recently, the radiative corrections to the kernel of the
 BFKL equation  had been calculated~\cite{LF89}~-~\cite{CCH} and the kernel
 for the forward scattering is presently known in the next-to-leading order (NLO)~\cite{FL98,CC98}.
 Attempts to apply it for the description of
 experiment \cite{Kim} are encouraging, but for a consistent comparison
 with the data one needs to know
 the impact factors  of colliding
 particles with the same accuracy as the  kernel of the BFKL equation.

 The cross section of interaction of particles $A$ and $B$
 is given in
 the BFKL approach by the convolution of the  Green's function $G$ with the
 impact factors $\Phi _{A}$ and $\Phi _{B}$.
 The  Green's function describes the propagation of two
 interacting
 Reggeized gluons, it is determined by the kernel of the BFKL equation.
 Whereas the impact factors  describe the interaction of
 the colliding particles with the Reggeized gluons.
 \begin{equation}
 \sigma_{AB}(s)=\int_{\delta-i\infty }^{\delta+i\infty
 }\frac{d\omega }{2\pi i}\, \int \frac{d^2q_A}{2\pi \vec q^{~2}_A}
 \int \frac{d^2q_B}{2\pi \vec q^{~2}_B}\left(\frac
 s{s_0}\right)^\omega \, \Phi _A({\vec q}_A,s_0) G_{\omega}({\vec
 q}_{A}, -{\vec q}_{B}) \Phi _B({\vec q}_B,s_0). \label{z1}
 \end{equation}
 Here the vector sign is used for vector components transverse to
 the plane of the initial momenta $p_A, p_B$, $G_\omega$ is the Mellin
 transform of the Green's function, and $s_0$ is an appropriately chosen
 energy scale. The representation (\ref{z1}) is valid both in the LLA and in the NLA.
 In both cases $ G_{\omega}({\vec
 q}_{A}, -{\vec q}_{B})$ is scale-independent. The impact factors become
 dependent in the NLA on the energy scale in such a way that under variations of $s_0$
 the cross section remains, within the NLA accuracy, unchanged.
 It was shown in \cite{F98} that to the NLA accuracy one can  change the scale
 $s_0$ in (\ref{z1}) for any factorizable scale $f_Af_B$, with $f_i$ depending on  ${\vec q}_i$,
 without changing the
 Green's function, provided that the impact factors are also changed correspondingly.

 At present impact factors are known in the NLO for the scattering of elementary coloured
 particles, quarks and gluons \cite{FFKP00,CR00}.
 In the case of the colorless particles scattering the corresponding results for
 the impact factors have been obtained in the leading order (LO) only.
 This restricts much a predictive power of the BFKL approach.
 In \cite{Kim} the phenomenological analysis is performed with use of the LO
 $\gamma^*$ impact factors\footnote{These impact factors differ by only trivial
 coefficients from the analogous ones in QED, which were obtained many years ago
 \cite{Cheng}.},
 therefore,  the theoretical predictions have some spread related to the choice of the
 energy scale $s_0$. To obtain  more certain predictions one needs to know the NLO
 $\gamma^*$ impact factors, that makes their evaluation a very important and timely problem.

 The knowledge of the NLO  $\Phi_{\gamma^*}$ is necessary not only at
 energies at which the BFKL dynamics is completely developed, but
 also in the case when only a few  terms of BFKL series  do
 contribute (that is  probably the case at modern energies). In this
 situation the NLO  $\Phi_{\gamma^*}$ determines a size of radiative
 corrections to the non-decreasing with $s$ contribution to the
 total cross section.

 The calculation of $\Phi_{\gamma^*}$ in the NLO have been
 started in \cite{FIK1}~-~\cite{22a}.
 The impact factors  are unambiguously defined \cite{FF98} in terms of the effective
 vertices for the Reggeon-particle interactions.  For the evaluation of $\Phi_{\gamma^*}$ one
 needs to know the amplitude $\Gamma _{\gamma^*R\rightarrow q\bar q}^{c}$ for the
 $q\bar q$- pair production in $\gamma^*$-Reggeon collision up to
 the one-loop accuracy as well as the amplitude at Born level describing the radiation of
 an additional gluon,
 $\Gamma _{\gamma^*R\rightarrow q\bar q g}^{c}$. They presently are
 both known~\cite{BGQ}~-~\cite{BGK} and can be used  for the calculation of $\Phi_{\gamma^*}$
 in the NLO. For this purpose the amplitudes must be squared and integrated over
 the phase space and the squared invariant mass of the produced particles.
 Unfortunately the corresponding expressions are extremely
 complicated which makes it difficult to progress further with analytical calculations.

 In this situation it is natural to look for an alternative possibility to
 organize evaluation of $\Phi_{\gamma^*}$.
 In this paper we suggest an
 approach based on intensive use of analytical properties of Feynman diagrams
 of an effective field theory with the Reggeized gluon. This
 method allows to consider part of the diagrams without their real
 calculation which simplifies considerably the problem.

 The paper is
 organized as follows. In the next section we give a diagrammatic 
 representation
 for the virtual photon impact factor. In the Sections~3~and~4 the analytical
 properties of the Feynman diagrams are discussed in length and the simplifications
 due to the analyticity are derived. The last section contains our conclusions.

 \section{Representation of the impact factor}
 \setcounter{equation}{0}

 We use the Sudakov decomposition of the virtual photon and
 the Reggeon momenta $p_A$ and $q$
 \begin{equation}\label{21}
 p_A = p_1 - \frac{Q^2}{s}p_2~,\ \ \ p_A^2 = -Q^2~,\ \ \ q = \alpha_Rp_2
 + q_\perp = \frac{\tilde s + Q^2 + \vec q^{~2}}{s}p_2 + q_\perp~,\ \ \
 q^2 = q_\perp^2 = - \vec q^{~2}~,
 \end{equation}
 where $(p_1, p_2)$ is the light-cone basis of the initial particles momenta
 plane
 \begin{equation}\label{22}
 p_1^2 = p_2^2 = 0~,\ \ \ 2p_1p_2 = s \rightarrow \infty~,
 \end{equation}
 and $\tilde s = (p_A + q)^2$ is the virtual photon - Reggeon invariant
 mass squared. It is of order of typical transverse momenta, i.e., limited and does
 not grow with $s$ since the impact factor is related by the definition to production
 of $q\bar q$ and $q\bar q g$ states in  the $\gamma^*$ fragmentation
 region.\footnote{The component of the Reggeon momentum proportional to Sudakov vector
 $p_1$ is irrelevant for the  analysis of the impact factor, therefore it is
 not included in the Eq.~(\ref{21}).}

 The impact factor $\Phi _{\gamma^*}$ in the LO, where a Reggeized gluon  acts
 as the ordinary gluon with polarization vector $p_2/s$, reads
 \begin{equation}
 \Phi _{\gamma^*}^{(0)}(\vec q)\;=\;\frac
 1{\sqrt{N_c^2-1}}\:\sum_{\{a\}}\:\int \:\frac{d \tilde s}{2\pi
 }\:|\Gamma _{\gamma^*R\rightarrow q\bar q}^{(0)c}|^2\:d\rho
 _{q\bar q} \ . \label{z2}
 \end{equation}
 Here $N_c=3$ is the number of QCD colours,
 $\Gamma_{\gamma^*R\rightarrow q\bar q}^{(0)c}$  is the amplitude of
 $q\bar q$ pair production in the $\gamma^*R$ collision, evaluated in
 the Born approximation,  $\tilde s$ is the squared c.m.s. energy of the
 $\gamma^* R$ system (equal in this case to the invariant mass  of the produced
 $q \bar q$  pair), $d\rho_{q\bar q}$ is the phase space element of the pair,
 see Ref.~\cite{FM98}.  The sum $\{a\}$ is over all discrete quantum numbers of the produced pair.

 The representation of $\Phi _{\gamma^*}$ in the NLO is more complicated.
 First of all, $\Phi_{\gamma^*}$ is expressed
 in terms of $\Gamma_{\gamma^*R\rightarrow f}^{c}$ in a more
 complicated way \cite{FF98} and depends on the energy scale $s_0$
 in the Mellin transform (\ref{z1}):
 \[
 \Phi_{\gamma^*}(\vec q, s_0)\;=\;\frac
 1{\sqrt{N_c^2-1}}\:\sum_{\{f\}}\:\int\:\frac{d\tilde s}{2\pi}\:
 |\Gamma_{\gamma^*R \rightarrow f}^{c}|^2\:d\rho_f\:\theta
 ({M^2-\tilde s})
 \]
 \begin{equation}
 -\frac{g^2N_c\vec q^{~2}}{(2\pi)^{D-1}}\int\frac{d^{D-2}q_1}{\vec
 q_1^{~2}(\vec q_1-\vec q)^2}\Phi^{(0)}_{\gamma^*}(\vec q_1)\;\ln
 \frac{M^4}{s_0(\vec q_1-\vec q)^2}-\omega^{(1)}(q^2)\Phi^{(0)}_{\gamma^*}(\vec q)\;\ln\frac{\vec q^{~2}}{s_0}\,
 , \label{z3}
 \end{equation}
 where the sum $\{f\}$ is over all discrete quantum numbers of the
 contributing intermediate states, which are now $q\bar q$ and $q\bar q g$;
 $M \rightarrow \infty$ is the cut off, which becomes
 necessary since the integral over $\tilde s$ is divergent for the
 ${q\bar q g}$ state. The dependence on $M$ in the  r.h.s of  Eq.~(\ref{z3})
 vanishes due to
 cancellation between the first and the second terms.
 $D=4+2\epsilon$ is the space-time dimension taken different from $D=4$
 for regularization of infrared and ultraviolet divergences which appear at the intermediate steps.
 Note that Eq.~(\ref{z3}) provides
 the infrared finiteness of the  impact factor as
 it was shown explicitly in Ref.~\cite{FM98}. The last two terms in Eq.~(\ref{z3})
 are to subtract the contribution coming from
 the emission of a gluon (both the real and the virtual one) outside the fragmentation region
 of the photon $\gamma^*$, the effects which were taken into account already in the LLA.
 \begin{equation}
 \omega^{(1)}(t)= -\frac{g^2N_c\Gamma(1-\epsilon)\Gamma^2(\epsilon)}{(4\pi)^{2+\epsilon}
 \Gamma(2\epsilon)}{(\vec q^{~2})^\epsilon}
 \label{z4}
 \end{equation}
 is the gluon Regge trajectory in the one-loop approximation.
 In the NLO the  Reggeon vertices depend on a scale of energy
 used in the Reggeon factor (see Ref.~\cite{Rio98} for more details).
 In the vertices $\Gamma_{\gamma^*R \rightarrow q\bar q}^{c}$ entering in
 Eq.~(\ref{z3})
 this scale is taken equal  $\vec q^{~2}$.

 The second complication is that in the NLO the Reggeon differs essentially
 from the gluon. Therefore, unlike usual QCD vertices (such as, for example,
 the quark-quark-gluon vertex) for which one can draw a definite set of Feynman diagrams
 with perfectly defined rules for calculation of their
 contributions, we have not such rules for the Reggeon vertices.
 Usually these vertices are extracted from a comparison of appropriate scattering amplitudes
 with their Reggeized form, so
 that to obtain a NLO  vertex one has to calculate radiative corrections to a whole amplitude.
 Nevertheless, it was shown in Refs.~\cite{FM98},~\cite{FF01}, that it is possible to formulate
 definite rules for a calculation of the vertices themselves. In the NLO any vertex can be presented
 as the sum of two contributions, the first of which is equal to
 $\Gamma^{(0)}\omega^{(1)}\left( q^2 \right)
 ( \ln\beta_0+k/2)$ where $\Gamma^{(0)}$ is
 the LO vertex, $\beta_0$ is an  intermediate parameter
 for separation of two kinematical regions, $k$ is the process independent constant
 \begin{equation}\label{26}
 k = \frac{1}{\epsilon} + 2\psi(1+2\epsilon) -
 2\psi(1+\epsilon) + \psi(1-\epsilon) - \psi(1)  + \frac{5 + 3\epsilon - 2(1+\epsilon)n_f/N_c}{2(1+2\epsilon)
 (3+2\epsilon)}~,
 \end{equation}
 and the second contribution  can be obtained as usual one-loop QCD amplitude with
 ordinary representation in terms of Feynman diagrams, where for the
 Reggeized gluon interactions one should use the vertices of  Fig.~1.

 \begin{figure}[tb]
 \begin{picture}(300,170)(0,-70)
 \ArrowLine(10,80)(50,50)\ArrowLine(50,50)(90,80)
 \ZigZag(50,0)(50,47){3}{4}\ArrowLine(50,47)(50,50)

 \Text(55,0)[l]{$q, c$}\Text(100,40)[l]{$=igt^c\frac{\not p_2}{s}~.$}
 \Text(100,-20)[c]{$(a)$}

 \Gluon(160,80)(196,53){3}{4}\Gluon(240,80)(204,53){3}{4}
 \ArrowLine(196,53)(200,50)\ArrowLine(204,53)(200,50)
 \ZigZag(200,0)(200,47){3}{4}\ArrowLine(200,47)(200,50)

 \Text(205,0)[l]{$q, c$}\Text(170,80)[l]{$k_1, a, \mu$}
 \Text(245,80)[l]{$k_2, b, \nu $} \Text(250,40)[l]{$=-igT^c_{ab}
 \biggl[g^{\mu\nu}\frac{p_2(k_2-k_1)}{s} + \frac{p_2^\mu}{s}(2k_1+k_2)^\nu $}
 \Text(50,-50)[l]{$-\frac{p_2^\nu}{s}(2k_2+k_1)^\mu + \frac{2q^2}{s}\frac
 {p_2^\mu p_2^\nu}{p_2(k_1-k_2)} \biggr]\theta\left(\left| \frac{p_2(k_1-
 k_2)}{s} \right| - \beta_0 \right)~.$}
 \Text(250,-20)[c]{$(b)$}
 \end{picture}

 \caption[] {The quark-quark-Reggeon and the gluon-gluon-Reggeon
 effective vertices. The zig-zag lines represent the Reggeized gluon;
 $t^c$ and $T^c$ are the colour group generators
 in the fundamental and adjoint representations respectively.}

 \end{figure}

 Self-energy insertions in the Reggeon line must be omitted and the Feynman
 gauge for  virtual gluons is assumed. In (\ref{26})
 $\psi(z)=\Gamma^\prime (z)/\Gamma(z)$,  $n_f$ is
 the number of quark flavours.  The intermediate parameter $\beta_0 \rightarrow 0$
 (remind, however, that first of all  the limit $s \rightarrow \infty$ must be taken)
 and the dependence on $\beta_0$ vanishes in this limit due to the
 cancellation between the two contributions.
 The  amplitude $\Gamma^c_{\gamma^* R\rightarrow q\bar qg}$  is needed at the Born level only.
 It can be calculated using the same Feynman rules (the vertices of Fig.~1 and
 the Feynman gauge for virtual gluons propagators). Generally speaking, for
 the case of real emission
 the theta function in the vertex of Fig.~1 should be absent. But for the real emission
 the limitation on the invariant mass of the
 $q\bar q g$ system in (\ref{z3}) leads to
 \begin{equation}\label{28}
 \left|\frac{p_2(k_1-k_2)}{s}\right|\geq \frac{\vec k^{~2}}{M^2}~,
 \end{equation}
 where $\vec k$ is the transverse momentum of the emitted gluon.
 We will  choose the parameter $\beta_0\rightarrow 0$ so small that
 in the region (\ref{28}) the theta function of Fig.~1 is equal 1, so
 that it can be formally written.
 The  amplitude obtained in such a way  is gauge invariant with respect
 to the emitted gluon, that allows to use the Feynman gauge for summation over its
 polarizations as well.

 Therefore we can present the NLO virtual photon impact factor in the form
 $$
 \Phi_{\gamma^*}(\vec q, s_0) = \Phi_M(\vec q)-\frac{g^2N_c\vec q^{~2}}{(2\pi)^{D-1}}\int\frac{d^{D-2}q_1}{\vec
 q_1^{~2}(\vec q_1-\vec q)^2}\Phi^{(0)}_{\gamma^*}(\vec q_1)\;\ln
 \frac{M^4}{s_0(\vec q_1-\vec q)^2}
 $$
 \begin{equation}\label{29}
 +\omega^{(1)}(q^2)\Phi^{(0)}_{\gamma^*}(\vec q)\;\left(\ln\frac{s_0\beta_0^2}{\vec q^{~2}}+k\right)\,,
 \end{equation}
 where $k$ is defined in Eq.~(\ref{26}) and  $\Phi_M$ is expressed
 in terms of the $\tilde s$-channel discontinuity of the  forward $\gamma^*R$ scattering amplitude:
 \begin{equation}\label{210}
 \Phi_M(\vec q) = -2i\int_0^{M^2}\frac{d\tilde s}{2\pi}\sum_n D_n^{(cut)}(q) =
 -2i\int_0^{M^2}\frac{d\tilde s}{2\pi}\Delta_{\tilde s}\sum_l D_l(q)~.
 \end{equation}
 Here $D_l$ is the contribution of $l$-th diagram to the
 $\gamma^*R \rightarrow \gamma^*R$ amplitude, $\Delta_{\tilde s} D_l$ is its $\tilde s$-
 channel discontinuity and $D_n^{(cut)}$ is the contribution to the discontinuity of $n$-th
 cut diagram calculated according to Cutcosky rule
 ($-2\pi i\delta_{+} (k^2)$ instead of  $1/(k^2+i0)$ for a cut
 line with momentum $k$). The factor $2$ in (\ref{210}) appears since each diagram has
 its partner with opposite direction of quark line, which  gives the same contribution,
 and which is not included in the sums in~(\ref{210}). There are $34$ LO and NLO diagrams
 for $\gamma^*R$ scattering having a discontinuity in the $\tilde
 s$-channel, and $56$ cut diagrams (some
 of diagrams have more than one Cutcosky cuts) in the
 relation~(\ref{210}) for the contribution $\Phi_M(\vec q)$ to the virtual photon impact factor,
 which we will be mostly concentrated on in the following. We
 do not present all of them here, but explain our method how to reduce the
 number of diagrams to be calculated on an example of the simple subset of
 diagrams shown in Fig.~2.
 In spite of the simplicity it contains all the relevant to our discussion
 features of the complete set of diagrams contributing to Eq.~(\ref{210}).
 \begin{figure}[tb]
 \begin{center}
 \begin{picture}(350,150)(0,-50)

 \Photon(0,50)(37,50){3}{4}\ArrowLine(37,50)(40,50)
 \Photon(110,50)(147,50){3}{4}\ArrowLine(147,50)(150,50)
 \BCirc(75,50){35}\ArrowLine(73.5,85)(76.5,85)\ArrowLine(76.5,15)(73.5,15)
 \ZigZag(55,-20)(55,19){3}{5}\ArrowLine(55,19)(55,22)
 \ZigZag(95,78)(95,-17){3}{12}\ArrowLine(95,-17)(95,-20)
 \Text(10,60)[c]{$p_A$}\Text(140,60)[c]{$p_A$}
 \Text(35,-10)[c]{$q, c$}\Text(115,-10)[c]{$q, c^\prime$}
 \Text(75,-40)[c]{$(a)$}

 \Photon(200,50)(237,50){3}{4}\ArrowLine(237,50)(240,50)
 \Photon(310,50)(347,50){3}{4}\ArrowLine(347,50)(350,50)
 \BCirc(275,50){35}\ArrowLine(273.5,85)(276.5,85)\ArrowLine(276.5,15)(273.5,15)
 \ZigZag(255,-20)(255,19){3}{5}\ArrowLine(255,19)(255,22)
 \ZigZag(295,78)(295,-17){3}{12}\ArrowLine(295,-17)(295,-20)
 \Text(210,60)[c]{$p_A$}\Text(340,60)[c]{$p_A$}
 \Text(235,-10)[c]{$q, c$}\Text(315,-10)[c]{$q, c^\prime$}
 \Text(275,-40)[c]{$(b)$}

 \GlueArc(44,33)(12,-47,110){3}{4}\GlueArc(275,15)(12,10,170){3}{4}

 \end{picture}
 \end{center}
 \begin{center}
 \begin{picture}(350,150)(0,-50)

 \Photon(40,50)(3,50){3}{4}\ArrowLine(3,50)(0,50)
 \Photon(150,50)(113,50){3}{4}\ArrowLine(113,50)(110,50)
 \BCirc(75,50){35}\ArrowLine(73.5,85)(76.5,85)\ArrowLine(76.5,15)(73.5,15)
 \ZigZag(55,22)(55,-17){3}{5}\ArrowLine(55,-17)(55,-20)
 \ZigZag(95,-20)(95,75){3}{12}\ArrowLine(95,75)(95,78)
 \Text(10,60)[c]{$p_A$}\Text(140,60)[c]{$p_A$}
 \Text(35,-10)[c]{$q, c^\prime$}\Text(115,-10)[c]{$q, c$}
 \Text(75,-40)[c]{$(c)$}

 \Photon(240,50)(203,50){3}{4}\ArrowLine(203,50)(200,50)
 \Photon(350,50)(313,50){3}{4}\ArrowLine(313,50)(310,50)
 \BCirc(275,50){35}\ArrowLine(273.5,85)(276.5,85)\ArrowLine(276.5,15)(273.5,15)
 \ZigZag(255,22)(255,-17){3}{5}\ArrowLine(255,-17)(255,-20)
 \ZigZag(295,-20)(295,75){3}{12}\ArrowLine(295,75)(295,78)
 \Text(210,60)[c]{$p_A$}\Text(340,60)[c]{$p_A$}
 \Text(235,-10)[c]{$q, c^\prime$}\Text(315,-10)[c]{$q, c$}
 \Text(275,-40)[c]{$(d)$}

 \GlueArc(44,33)(12,-47,110){3}{4}\GlueArc(275,15)(12,10,170){3}{4}

 \end{picture}
 \end{center}

 \caption[]{Some of  diagrams contributing to  Eq.~(\ref{210}).}

 \end{figure}
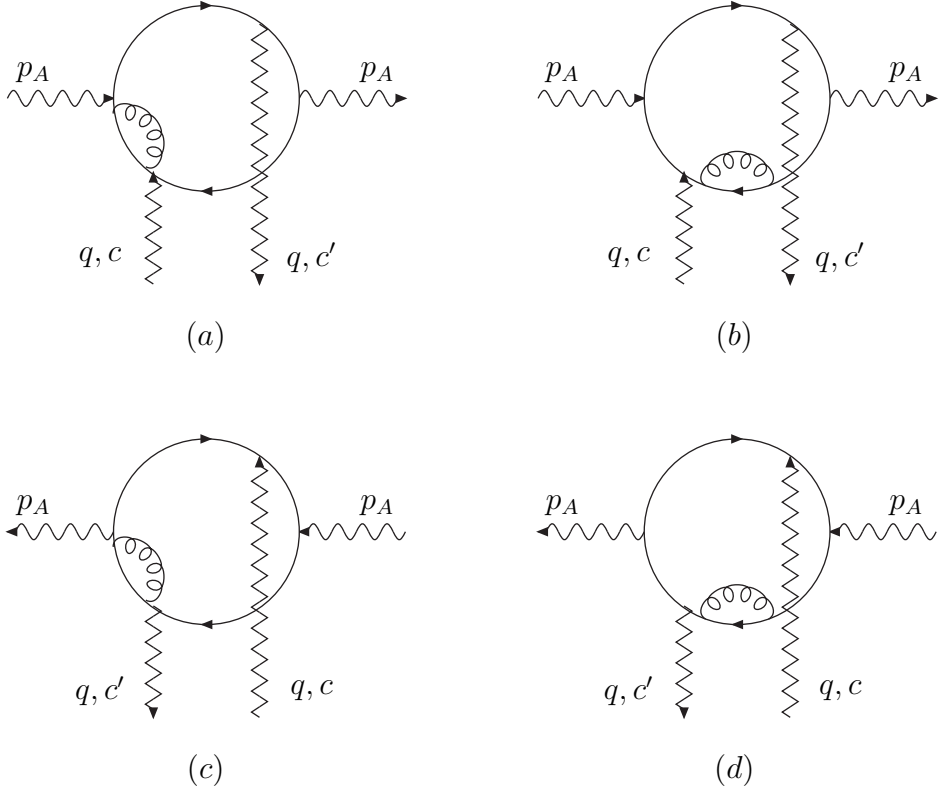

 First we notice that the diagrams $(a)$ and $(c)$ in Fig.~2, as well as
 the diagrams $(b)$ and $(d)$ there, differ by the directions of all  external
 lines only. The replacement $c \leftrightarrow c^\prime$ does not change
 anything because of the colour singlet in the $t$- channel. This means that
 we may consider, for instance, the first two diagrams of Fig.~2
 only, assuming for them
 \begin{equation}\label{211}
 e_\mu e^{\prime*}_\nu \rightarrow e_\mu
 e^{\prime*}_\nu + e^{\prime*}_\mu e_\nu~,
 \end{equation}
 with $e$ and $e^\prime$ being the polarizations vectors of the incoming and outgoing
 virtual photons correspondingly. Each diagram in  Eq.~(\ref{210})
 either has such a partner or it is self-symmetric under the symmetrization of
 the initial and final virtual photons polarizations. Further we always
 assume the prescription~(\ref{211}) for each diagram we work with, that
 reduces the number of Feynman diagrams to be considered in the
 relation~(\ref{210}) to 22, 2 (LO) and 20 (NLO) shown in Figs.~4,~5(a) and
 Fig.~7. In the next sections we will show how to reduce
 more this number using analytical properties of the diagrams.

 \section{Analytical properties of $D_n$}
 \setcounter{equation}{0}

 Let us consider one of the contributions  $D_n$ in
 Eq.~(\ref{210}). It is an analytical function of variable $\tilde s$ having
 a branch cut discontinuity
 at $0 < \tilde s < \infty$ and, possibly, also
 at $-\infty < \tilde s < -2\left( \vec q^{~2} + Q^2 \right)$ related to
 the $\tilde u$- channel Cutcosky cuts with
 \begin{equation}\label{31}
 \tilde u=(p_A - q)^2 = - 2\left( \vec q^{~2} + Q^2 \right) - \tilde s~.
 \end{equation}
 We need the integral (see Fig.~3)

 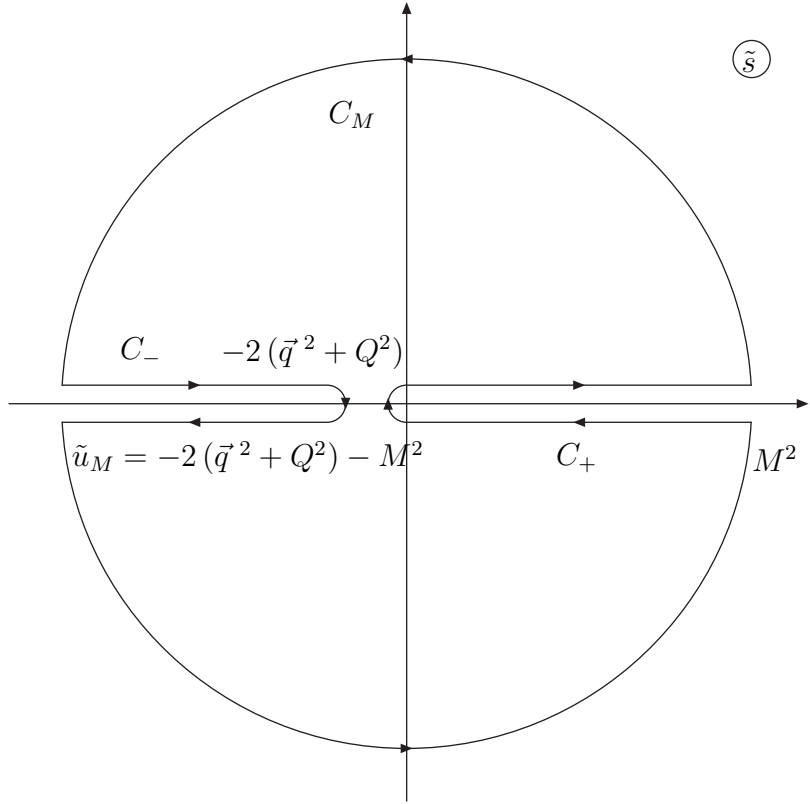
\begin{figure}[tb]

 \begin{center}
 \begin{picture}(300,300)(0,0)

 \LongArrow(0,150)(300,150)\LongArrow(150,0)(150,300)
 \ArrowArc(150,150)(130,3,177)\ArrowArc(150,150)(130,183,357)
 \ArrowLine(150,157)(280,157)\ArrowLine(280,143)(150,143)
 \ArrowArcn(150,150)(7,270,90)
 \ArrowLine(20,157)(120,157)\ArrowLine(120,143)(20,143)
 \ArrowArcn(120,150)(7,90,270)
 \BCirc(280,280){7}\Text(280,280)[]{$\tilde s$}
 \Text(290,130)[]{$M^2$}\Text(215,130)[]{$C_+$}
 \Text(115,170)[]{$-2\left( \vec q^{~2} + Q^2 \right)$}
 \Text(50,170)[]{$C_-$}\Text(130,260)[]{$C_M$}
 \Text(24,130)[l]{$\tilde u_M = - 2\left( \vec q^{~2} + Q^2 \right)
 - M^2$}

 \end{picture}
 \end{center}

 \caption[]{Schematic representation of analytical
 properties of $D_n$.}

 \end{figure}

 \begin{equation}\label{32}
 \int_0^{M^2}d\tilde s\Delta_{\tilde s} D_n(q) = \int_{C_+}d\tilde s D_n(q) =
 - \int_{C_M}d\tilde s D_n(q) - \int_{C_-}d\tilde s D_n(q)~.
 \end{equation}
 Using Eqs.~(\ref{21}), (\ref{31}), the last integral here can be
 presented as
 \begin{equation}\label{33}
 - \int_{C_-}d\tilde s D_n(q) = \int_0^{M^2}d\tilde u\Delta_{\tilde u}D_n(q) =\int_0
 ^{M^2}d\tilde s\Delta_{\tilde s} D_n(-q^\prime) = \int_0^{M^2}
 d\tilde s\Delta_{\tilde s}\tilde D_n(q^\prime)~,
 \end{equation}
 where
 \begin{equation}\label{34}
 q^\prime = \frac{\tilde s + Q^2 + \vec q^{~2}}{s}p_2 - q_\perp = q\biggl|_
 {q_\perp \leftrightarrow - q_\perp}~,
 \end{equation}
 and  $\tilde D_n$ differs from  $D_n$ by the directions of Reggeon lines
 only. In the light-cone gauge
 \begin{equation}\label{35}
 ep_2 = e^\prime p_2 = 0~,
 \end{equation}
 which we  use, the contribution of each diagram can be decomposed over four
 independent spin structures which we define as follows
 $$
 T^{(+)} = \frac{-e_\mu e^{\prime*}_\nu}{1+\epsilon}g_{\perp}^{\mu\nu}
 ~,\ \ \ T^{(-)} = \frac{-e_\mu e^{\prime*}_\nu}{(1+\epsilon)(1+2\epsilon)}
 \left( g_{\perp}^{\mu\nu} - (D-2)\frac{q_\perp^\mu q_\perp^\nu}{q_\perp
 ^2} \right)~,
 $$
 \begin{equation}\label{36}
 L^{(+)} = e_\mu e^{\prime*}_\nu\frac{p_A^\mu p_A^\nu}{Q^2}~,\ \ \ L^{(-)} =
 \left( e_\mu e^{\prime*}_\nu + e^{\prime*}_\mu e_\nu \right)\frac{2p_A^\mu
 q_\perp^\nu}{\sqrt{2\vec q^{~2}Q^2}}~,
 \end{equation}
 where
 \begin{equation}\label{36a}
 g_{\perp}^{\mu\nu} = g^{\mu\nu} - \frac{p_1^\mu p_2^\nu + p_2^\mu p_1^\nu}{p_1p_2}~.
 \end{equation}
 The spin structures $T^{(+)}$ and $L^{(+)}$ describe the transverse and the longitudinal
 spin-non-flip transitions respectively, the other two structures, $T^{(-)}$ and $L^{(-)}$,
 correspond to the double and the single spin-flip helicity amplitudes.
 Therefore we come to the conclusion
 \begin{equation}\label{37}
 \tilde D_n(q^\prime) = \tilde D_n(q)\biggl|_{q_\perp \leftrightarrow - q_\perp}
 = \tilde D_n(q)\biggl|_{L^{(-)} \leftrightarrow - L^{(-)}}~,
 \end{equation}
 and finally, using Eqs.~(\ref{32}), (\ref{33}) and~(\ref{37}), we get
 \begin{equation}\label{38}
 \int_0^{M^2}d\tilde s\Delta_{\tilde s} D_n(q) = - \int_{C_M}d\tilde s D_n(q)
 + \int_0^{M^2}d\tilde s\Delta_{\tilde s}\tilde D_n(q)\biggl|_{L^{(-)}
 \leftrightarrow - L^{(-)}}~.
 \end{equation}

 Eq.~(\ref{38}) can be written for any diagram in~(\ref
 {210});  for those of them without $\tilde u$- channel
 cut the last term in ~(\ref{38}) does not appear. Let us consider the application of
 the relation~(\ref{38}) for the diagrams in  Fig.~2.
 According to the discussion above, their contribution to $\Phi_M$
 takes the form
 $$
 \Phi_M^{(Fig.~2)} = - 2i\int_0^{M^2}\frac{d\tilde s}{2\pi}\Delta
 _{\tilde s}\left( D_{2(a)} + D_{2(b)} \right)
 $$
 \begin{equation}\label{39}
 = 2i\int_{C_M}\frac{d\tilde s}{2\pi}D_{2(b)} - 2i\int_0^{M^2}\frac
 {d\tilde s}{2\pi}\Delta_{\tilde s}\left( D_{2(a)} + \tilde D_{2(b)}\biggl|_{L^{(-)}
 \leftrightarrow - L^{(-)}} \right)~,
 \end{equation}
 where we have applied~(\ref{38}) to the diagram $D_{2(b)}$. It is easy to see that
 \begin{equation}\label{310}
 \tilde D_{2(b)} = D_{2(a)}~,
 \end{equation}
 therefore
 \begin{equation}\label{311}
 \Phi_M^{(Fig.~2)} = 2i\int_{C_M}\frac{d\tilde s}{2\pi}D_{2(b)} -
 4i\int_0^{M^2}\frac{d\tilde s}{2\pi}\Delta_{\tilde s} D_{2(a)}\biggl|_{L^{(-)}
 = 0}~.
 \end{equation}

 Comparing  Eqs.~(\ref{39}) and~(\ref{311}), we see, that if we worked in a
 scalar quantum field theory (let us say $\phi^3$ for definiteness) we could
 consider only one diagram instead of  two, because the contribution of the
 large circle
 would disappear due to  fast enough decrease of
 amplitudes at large $\tilde s$. The situation in QCD is more complicated:
 contributions from the integration over the large  circle  basically
 survive for separate
 diagrams. Nevertheless, we will show in the following that
 analyticity helps to consider many diagrams without their real
 calculation in the QCD case also.

 After a short consideration it becomes clear
 that the transformations we used to obtain  (\ref{311}), the result for the
 contribution of the set of diagrams shown in Fig.~2, may be applied as well to the
 complete set of 22 diagrams which contribute to Eq.~(\ref{210}). Proceeding
 along the same steps we arrive at the following representation of the impact factor
 \begin{equation}\label{312}
 \Phi_M = \Phi_\Delta + \Phi_\Lambda~,
 \end{equation}
 where the first term is given by the integral over $\tilde s$-
 channel discontinuity of 12 diagrams of Fig.~4,
  $$
 \Phi_\Delta = -i\int_0^{M^2}\frac{d\tilde s}{2\pi}\Delta_{\tilde s}\biggl[
 D_{4(a)} + \cdots + D_{4(c)}
 $$
 \begin{equation}\label{313}
 + 2\left( D_{4(d)} + \cdots + D_{4(j)} \right) + 2\left( D_{4(k)} +
 2D_{4(l)}
 \right)\biggl|_{L^{(-)} = 0} \biggr]~,
 \end{equation}
 whereas
 the second contribution, $\Phi_\Lambda$,  consists of the
 integrals over the large circle from 10 diagrams
 shown in Fig.~5(a) and in Fig.~7.
 To see this one needs to move the integral over the $\tilde s$- channel
 discontinuity for these diagrams to the integral over the infinite circle.
 Two of the diagrams in Fig.~7, $D^r_2$ and $D^r_3$, have
 both $\tilde s$- and $\tilde u$- channel singularities.
 As it was demonstrated above
 the contributions of the $\tilde u$- channel cuts for these diagrams may be
 related to the integrals over $\tilde s$- channel discontinuities of their
 symmetric partners, diagrams $(l)$ and $(k)$ in Fig.~4.
 That is quite natural since the $\tilde s$- and $\tilde u$- channels
 are actually the same in the case under consideration.
 We used this symmetry deriving our representation for the impact factor,
 this is why factor 2 is present in the last term of Eq.~(\ref{313}).
 The discontinuity $\Delta_{\tilde s}$ in this relation is given by
 the  $\tilde s$- channel Cutcosky cuts. There are 18 cut diagrams corresponding
 to the diagrams of Fig.~4.

 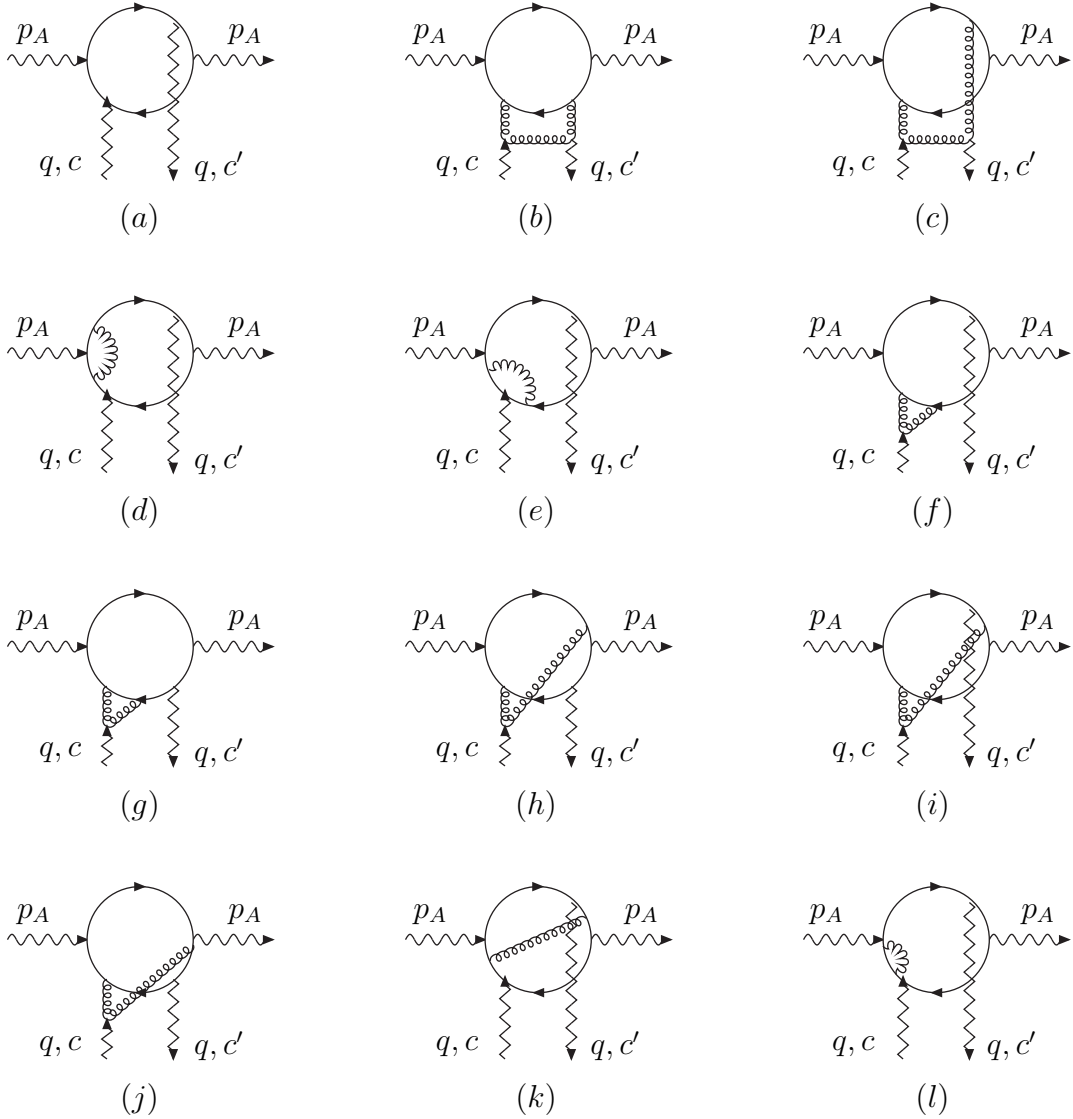
\begin{figure}[tb]
 \vspace{-1.7cm}

 \begin{center}
 \begin{picture}(400,110)(0,0)

 \Photon(0,50)(27,50){2}{4}\ArrowLine(27,50)(30,50)
 \Photon(70,50)(97,50){2}{4}\ArrowLine(97,50)(100,50)
 \BCirc(50,50){20}\ArrowLine(48.5,70)(51.5,70)\ArrowLine(51.5,30)(48.5,30)
 \ZigZag(37.5,5)(37.5,33){2}{4}\ArrowLine(37.5,33)(37.5,36)
 \ZigZag(62.5,64)(62.5,8){2}{8}\ArrowLine(62.5,8)(62.5,5)
 \Text(10,60)[]{$p_A$}\Text(90,60)[]{$p_A$}\Text(20,10)[]{$q, c$}
 \Text(80,10)[]{$q, c^\prime$}\Text(50,-10)[]{$(a)$}

 \Photon(150,50)(177,50){2}{4}\ArrowLine(177,50)(180,50)
 \Photon(220,50)(247,50){2}{4}\ArrowLine(247,50)(250,50)
 \BCirc(200,50){20}\ArrowLine(198.5,70)(201.5,70)\ArrowLine(201.5,30)(198.5,30)
 \ZigZag(187.5,5)(187.5,17){2}{2}\ArrowLine(187.5,17)(187.5,20)
 \ZigZag(212.5,20)(212.5,8){2}{2}\ArrowLine(212.5,8)(212.5,5)
 \Text(160,60)[]{$p_A$}\Text(240,60)[]{$p_A$}\Text(170,10)[]{$q, c$}
 \Text(230,10)[]{$q, c^\prime$}\Text(200,-10)[]{$(b)$}
 \Gluon(187.5,20)(187.5,35){1.5}{4}\Gluon(212.5,20)(212.5,35){-1.5}{4}
 \Gluon(187.5,20)(212.5,20){-1.5}{7}

 \Photon(300,50)(327,50){2}{4}\ArrowLine(327,50)(330,50)
 \Photon(370,50)(397,50){2}{4}\ArrowLine(397,50)(400,50)
 \BCirc(350,50){20}\ArrowLine(348.5,70)(351.5,70)\ArrowLine(351.5,30)(348.5,30)
 \ZigZag(337.5,5)(337.5,17){2}{2}\ArrowLine(337.5,17)(337.5,20)
 \ZigZag(362.5,20)(362.5,8){2}{2}\ArrowLine(362.5,8)(362.5,5)
 \Text(310,60)[]{$p_A$}\Text(390,60)[]{$p_A$}\Text(320,10)[]{$q, c$}
 \Text(380,10)[]{$q, c^\prime$}\Text(350,-10)[]{$(c)$}
 \Gluon(337.5,20)(337.5,35){1.5}{4}\Gluon(362.5,20)(362.5,65){-1.5}{12}
 \Gluon(337.5,20)(362.5,20){-1.5}{7}

 \end{picture}
 \begin{picture}(400,110)(0,0)

 \Photon(0,50)(27,50){2}{4}\ArrowLine(27,50)(30,50)
 \Photon(70,50)(97,50){2}{4}\ArrowLine(97,50)(100,50)
 \BCirc(50,50){20}\ArrowLine(48.5,70)(51.5,70)\ArrowLine(51.5,30)(48.5,30)
 \ZigZag(37.5,5)(37.5,33){2}{4}\ArrowLine(37.5,33)(37.5,36)
 \ZigZag(62.5,64)(62.5,8){2}{8}\ArrowLine(62.5,8)(62.5,5)
 \Text(10,60)[]{$p_A$}\Text(90,60)[]{$p_A$}\Text(20,10)[]{$q, c$}
 \Text(80,10)[]{$q, c^\prime$}\Text(50,-10)[]{$(d)$}
 \GlueArc(30,50)(10,75,-75){-1.5}{8}

 \Photon(150,50)(177,50){2}{4}\ArrowLine(177,50)(180,50)
 \Photon(220,50)(247,50){2}{4}\ArrowLine(247,50)(250,50)
 \BCirc(200,50){20}\ArrowLine(198.5,70)(201.5,70)\ArrowLine(201.5,30)(198.5,30)
 \ZigZag(187.5,5)(187.5,33){2}{4}\ArrowLine(187.5,33)(187.5,36)
 \ZigZag(212.5,64)(212.5,8){2}{8}\ArrowLine(212.5,8)(212.5,5)
 \Text(160,60)[]{$p_A$}\Text(240,60)[]{$p_A$}\Text(170,10)[]{$q, c$}
 \Text(230,10)[]{$q, c^\prime$}\Text(200,-10)[]{$(e)$}
 \GlueArc(187.5,36)(10,127,-32){-1.5}{8}

 \Photon(300,50)(327,50){2}{4}\ArrowLine(327,50)(330,50)
 \Photon(370,50)(397,50){2}{4}\ArrowLine(397,50)(400,50)
 \BCirc(350,50){20}\ArrowLine(348.5,70)(351.5,70)\ArrowLine(351.5,30)(348.5,30)
 \ZigZag(337.5,5)(337.5,17){2}{2}\ArrowLine(337.5,17)(337.5,20)
 \ZigZag(362.5,64)(362.5,8){2}{8}\ArrowLine(362.5,8)(362.5,5)
 \Text(310,60)[]{$p_A$}\Text(390,60)[]{$p_A$}\Text(320,10)[]{$q, c$}
 \Text(380,10)[]{$q, c^\prime$}\Text(350,-10)[]{$(f)$}
 \Gluon(337.5,20)(337.5,35){1.5}{4}\Gluon(337.5,20)(350,30){-1.5}{4}

 \end{picture}
 \begin{picture}(400,110)(0,0)

 \Photon(0,50)(27,50){2}{4}\ArrowLine(27,50)(30,50)
 \Photon(70,50)(97,50){2}{4}\ArrowLine(97,50)(100,50)
 \BCirc(50,50){20}\ArrowLine(48.5,70)(51.5,70)\ArrowLine(51.5,30)(48.5,30)
 \ZigZag(37.5,5)(37.5,17){2}{2}\ArrowLine(37.5,17)(37.5,20)
 \ZigZag(62.5,35)(62.5,8){2}{4}\ArrowLine(62.5,8)(62.5,5)
 \Text(10,60)[]{$p_A$}\Text(90,60)[]{$p_A$}\Text(20,10)[]{$q, c$}
 \Text(80,10)[]{$q, c^\prime$}\Text(50,-10)[]{$(g)$}
 \Gluon(37.5,20)(37.5,35){1.5}{4}\Gluon(37.5,20)(50,30){-1.5}{4}

 \Photon(150,50)(177,50){2}{4}\ArrowLine(177,50)(180,50)
 \Photon(220,50)(247,50){2}{4}\ArrowLine(247,50)(250,50)
 \BCirc(200,50){20}\ArrowLine(198.5,70)(201.5,70)\ArrowLine(201.5,30)(198.5,30)
 \ZigZag(187.5,5)(187.5,17){2}{2}\ArrowLine(187.5,17)(187.5,20)
 \ZigZag(212.5,35)(212.5,8){2}{4}\ArrowLine(212.5,8)(212.5,5)
 \Text(160,60)[]{$p_A$}\Text(240,60)[]{$p_A$}\Text(170,10)[]{$q, c$}
 \Text(230,10)[]{$q, c^\prime$}\Text(200,-10)[]{$(h)$}
 \Gluon(187.5,20)(187.5,35){1.5}{4}\Gluon(187.5,20)(218,58){-1.5}{12}

 \Photon(300,50)(327,50){2}{4}\ArrowLine(327,50)(330,50)
 \Photon(370,50)(397,50){2}{4}\ArrowLine(397,50)(400,50)
 \BCirc(350,50){20}\ArrowLine(348.5,70)(351.5,70)\ArrowLine(351.5,30)(348.5,30)
 \ZigZag(337.5,5)(337.5,17){2}{2}\ArrowLine(337.5,17)(337.5,20)
 \ZigZag(362.5,64)(362.5,8){2}{8}\ArrowLine(362.5,8)(362.5,5)
 \Text(310,60)[]{$p_A$}\Text(390,60)[]{$p_A$}\Text(320,10)[]{$q, c$}
 \Text(380,10)[]{$q, c^\prime$}\Text(350,-10)[]{$(i)$}
 \Gluon(337.5,20)(337.5,35){1.5}{4}\Gluon(337.5,20)(368,58){-1.5}{12}

 \end{picture}
 \begin{picture}(400,110)(0,0)

 \Photon(0,50)(27,50){2}{4}\ArrowLine(27,50)(30,50)
 \Photon(70,50)(97,50){2}{4}\ArrowLine(97,50)(100,50)
 \BCirc(50,50){20}\ArrowLine(48.5,70)(51.5,70)\ArrowLine(51.5,30)(48.5,30)
 \ZigZag(37.5,5)(37.5,17){2}{2}\ArrowLine(37.5,17)(37.5,20)
 \ZigZag(62.5,35)(62.5,8){2}{4}\ArrowLine(62.5,8)(62.5,5)
 \Text(10,60)[]{$p_A$}\Text(90,60)[]{$p_A$}\Text(20,10)[]{$q, c$}
 \Text(80,10)[]{$q, c^\prime$}\Text(50,-10)[]{$(j)$}
 \Gluon(37.5,20)(37.5,35){1.5}{4}\Gluon(37.5,20)(70,48){-1.5}{12}

 \Photon(150,50)(177,50){2}{4}\ArrowLine(177,50)(180,50)
 \Photon(220,50)(247,50){2}{4}\ArrowLine(247,50)(250,50)
 \BCirc(200,50){20}\ArrowLine(198.5,70)(201.5,70)\ArrowLine(201.5,30)(198.5,30)
 \ZigZag(187.5,5)(187.5,33){2}{4}\ArrowLine(187.5,33)(187.5,36)
 \ZigZag(212.5,64)(212.5,8){2}{8}\ArrowLine(212.5,8)(212.5,5)
 \Text(160,60)[]{$p_A$}\Text(240,60)[]{$p_A$}\Text(170,10)[]{$q, c$}
 \Text(230,10)[]{$q, c^\prime$}\Text(200,-10)[]{$(k)$}
 \Gluon(182,42)(218,58){1.5}{11}

 \Photon(300,50)(327,50){2}{4}\ArrowLine(327,50)(330,50)
 \Photon(370,50)(397,50){2}{4}\ArrowLine(397,50)(400,50)
 \BCirc(350,50){20}\ArrowLine(348.5,70)(351.5,70)\ArrowLine(351.5,30)(348.5,30)
 \ZigZag(337.5,5)(337.5,33){2}{4}\ArrowLine(337.5,33)(337.5,36)
 \ZigZag(362.5,64)(362.5,8){2}{8}\ArrowLine(362.5,8)(362.5,5)
 \Text(310,60)[]{$p_A$}\Text(390,60)[]{$p_A$}\Text(320,10)[]{$q, c$}
 \Text(380,10)[]{$q, c^\prime$}\Text(350,-10)[]{$(l)$}
 \GlueArc(332,42)(6,110,-60){-1.5}{5}

 \end{picture}
 \end{center}

 \caption[]{The diagrams contributing to $\Phi_\Delta$.}

 \end{figure}

 In the next section we will concentrate on
 the  contribution $\Phi_\Lambda$ in  Eq.~(\ref{312}) given by the integrals over
 the large circle. We will prove that $\Phi_\Lambda$ does not depend on the
 Reggeon transverse momentum $\vec q$. That makes it possible
 to express $\Phi_M(\vec q)$
 through $\Phi_\Delta$~(\ref{313}) only, and
 to avoid, therefore, an explicit
 calculation of $\Phi_\Lambda$.

 \section{Large circle contribution}
 \setcounter{equation}{0}

 In order to consider the contribution $\Phi_\Lambda$ we introduce  new
 longitudinal subspace basis $(p_1, p_2^\prime)$ with
 \begin{equation}\label{41}
 p_2^\prime = \frac{\tilde s + Q^2 + \vec q^{~2}}{s}p_2~,\ \ \ 2p_1p_2^\prime = s_1
 = \tilde s + Q^2 + \vec q^{~2} = \frac{\tilde s - \tilde u}{2}~.
 \end{equation}
 Therefore
 \begin{equation}\label{42}
 q = p_2^\prime + q_\perp~,\ \ \ p_A = p_1 - \frac{Q^2}{s_1}p_2^\prime~,
 \end{equation}
 and the gauge fixing condition for the external virtual photons remains actually
 the same (compare with the Eq.(\ref{35}))
 \begin{equation}\label{43}
 ep_2^\prime = e^\prime p_2^\prime = 0~.
 \end{equation}

 Let us consider  first the Born contribution to $\Phi_\Lambda$
 \begin{equation}\label{44}
 \Phi_\Lambda^{(0)} = - i\left( eq_fg \right)^2\frac{\sqrt{N_c^2-1}}{4\pi}
 \int_{C_M}\frac{d\tilde s}{s_1^2}D^{r}_a~,
 \end{equation}
 where $eq_f$ is the quark electric charge and $D^{r}_a$ is the amplitude corresponding to  the diagram of Fig.~5(a),
 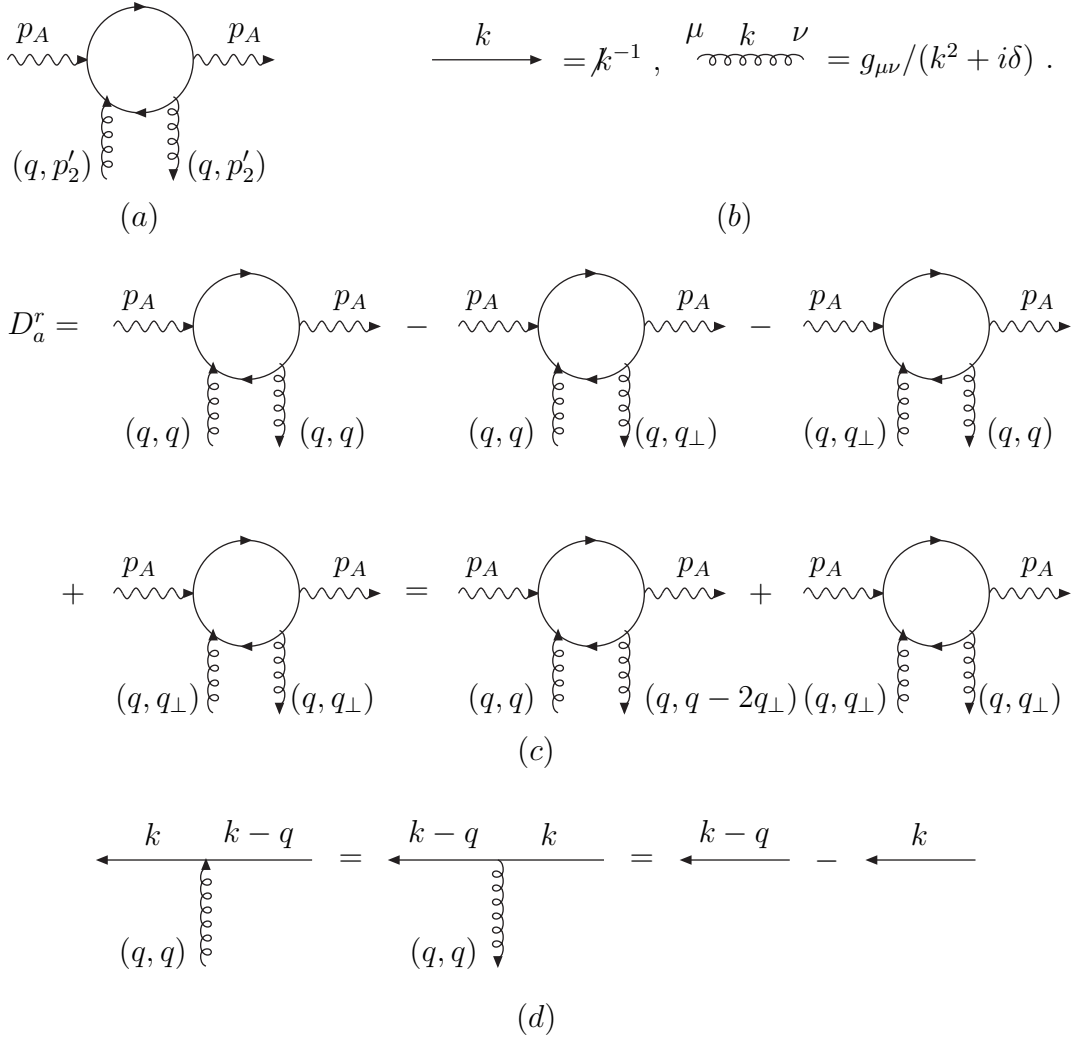
\begin{figure}[tb]
 \vspace{-1.7cm}

 \begin{center}
 \begin{picture}(400,100)(0,0)

 \Photon(0,50)(27,50){2}{4}\ArrowLine(27,50)(30,50)
 \Photon(70,50)(97,50){2}{4}\ArrowLine(97,50)(100,50)
 \BCirc(50,50){20}\ArrowLine(48.5,70)(51.5,70)\ArrowLine(51.5,30)(48.5,30)
 \Gluon(37.5,5)(37.5,33){2}{4}\ArrowLine(37.5,33)(37.5,36)
 \Gluon(62.5,36)(62.5,8){2}{4}\ArrowLine(62.5,8)(62.5,5)
 \Text(10,60)[]{$p_A$}\Text(90,60)[]{$p_A$}\Text(17,10)[]{$(q, p_2^\prime)$}
 \Text(83,10)[]{$(q, p_2^\prime)$}\Text(50,-10)[]{$(a)$}

 \LongArrow(160,50)(200,50)\Text(180,60)[]{$k$}\Text(210,50)[l]{$=\not k^{-1}~,$}
 \Gluon(260,50)(300,50){2}{6}\Text(260,60)[]{$\mu$}\Text(300,60)[]{$\nu$}
 \Text(280,60)[]{$k$}\Text(310,50)[l]{$=g_{\mu\nu}/(k^2 + i\delta)~.$}
 \Text(275,-10)[]{$(b)$}

 \end{picture}
 \begin{picture}(400,100)(0,0)

 \Photon(40,50)(67,50){2}{4}\ArrowLine(67,50)(70,50)
 \Photon(110,50)(137,50){2}{4}\ArrowLine(137,50)(140,50)
 \BCirc(90,50){20}\ArrowLine(88.5,70)(91.5,70)\ArrowLine(91.5,30)(88.5,30)
 \Gluon(77.5,5)(77.5,33){2}{4}\ArrowLine(77.5,33)(77.5,36)
 \Gluon(102.5,36)(102.5,8){2}{4}\ArrowLine(102.5,8)(102.5,5)
 \Text(50,60)[]{$p_A$}\Text(130,60)[]{$p_A$}\Text(57,10)[]{$(q, q)$}
 \Text(123,10)[]{$(q, q)$}\Text(0,50)[l]{$D^{r}_a=$}

 \Photon(170,50)(197,50){2}{4}\ArrowLine(197,50)(200,50)
 \Photon(240,50)(267,50){2}{4}\ArrowLine(267,50)(270,50)
 \BCirc(220,50){20}\ArrowLine(218.5,70)(221.5,70)\ArrowLine(221.5,30)(218.5,30)
 \Gluon(207.5,5)(207.5,33){2}{4}\ArrowLine(207.5,33)(207.5,36)
 \Gluon(232.5,36)(232.5,8){2}{4}\ArrowLine(232.5,8)(232.5,5)
 \Text(180,60)[]{$p_A$}\Text(260,60)[]{$p_A$}\Text(187,10)[]{$(q, q)$}
 \Text(253,10)[]{$(q, q_\perp)$}\Text(155,50)[]{$-$}

 \Photon(300,50)(327,50){2}{4}\ArrowLine(327,50)(330,50)
 \Photon(370,50)(397,50){2}{4}\ArrowLine(397,50)(400,50)
 \BCirc(350,50){20}\ArrowLine(348.5,70)(351.5,70)\ArrowLine(351.5,30)(348.5,30)
 \Gluon(337.5,5)(337.5,33){2}{4}\ArrowLine(337.5,33)(337.5,36)
 \Gluon(362.5,36)(362.5,8){2}{4}\ArrowLine(362.5,8)(362.5,5)
 \Text(310,60)[]{$p_A$}\Text(390,60)[]{$p_A$}\Text(317,10)[]{$(q, q_\perp)$}
 \Text(383,10)[]{$(q, q)$}\Text(285,50)[]{$-$}

 \end{picture}
 \begin{picture}(400,100)(0,0)

 \Photon(40,50)(67,50){2}{4}\ArrowLine(67,50)(70,50)
 \Photon(110,50)(137,50){2}{4}\ArrowLine(137,50)(140,50)
 \BCirc(90,50){20}\ArrowLine(88.5,70)(91.5,70)\ArrowLine(91.5,30)(88.5,30)
 \Gluon(77.5,5)(77.5,33){2}{4}\ArrowLine(77.5,33)(77.5,36)
 \Gluon(102.5,36)(102.5,8){2}{4}\ArrowLine(102.5,8)(102.5,5)
 \Text(50,60)[]{$p_A$}\Text(130,60)[]{$p_A$}\Text(57,10)[]{$(q, q_\perp)$}
 \Text(123,10)[]{$(q, q_\perp)$}\Text(25,50)[]{$+$}

 \Photon(170,50)(197,50){2}{4}\ArrowLine(197,50)(200,50)
 \Photon(240,50)(267,50){2}{4}\ArrowLine(267,50)(270,50)
 \BCirc(220,50){20}\ArrowLine(218.5,70)(221.5,70)\ArrowLine(221.5,30)(218.5,30)
 \Gluon(207.5,5)(207.5,33){2}{4}\ArrowLine(207.5,33)(207.5,36)
 \Gluon(232.5,36)(232.5,8){2}{4}\ArrowLine(232.5,8)(232.5,5)
 \Text(180,60)[]{$p_A$}\Text(260,60)[]{$p_A$}\Text(187,10)[]{$(q, q)$}
 \Text(270,10)[]{$(q, q-2q_\perp)$}\Text(155,50)[]{$=$}

 \Photon(300,50)(327,50){2}{4}\ArrowLine(327,50)(330,50)
 \Photon(370,50)(397,50){2}{4}\ArrowLine(397,50)(400,50)
 \BCirc(350,50){20}\ArrowLine(348.5,70)(351.5,70)\ArrowLine(351.5,30)(348.5,30)
 \Gluon(337.5,5)(337.5,33){2}{4}\ArrowLine(337.5,33)(337.5,36)
 \Gluon(362.5,36)(362.5,8){2}{4}\ArrowLine(362.5,8)(362.5,5)
 \Text(310,60)[]{$p_A$}\Text(390,60)[]{$p_A$}\Text(317,10)[]{$(q, q_\perp)$}
 \Text(383,10)[]{$(q, q_\perp)$}\Text(285,50)[]{$+$}

 \Text(200,-10)[]{$(c)$}

 \end{picture}
 \begin{picture}(330,100)(0,0)

 \LongArrow(80,50)(0,50)\Gluon(40,10)(40,47){2}{6}\ArrowLine(40,47)(40,50)
 \Text(20,60)[]{$k$}\Text(60,60)[]{$k-q$}\Text(20,15)[]{$(q, q)$}
 \Text(95,50)[]{$=$}

 \LongArrow(190,50)(110,50)\Gluon(150,50)(150,13){2}{6}\ArrowLine(150,13)(150,10)
 \Text(130,60)[]{$k-q$}\Text(170,60)[]{$k$}\Text(130,15)[]{$(q, q)$}
 \Text(205,50)[]{$=$}

 \LongArrow(260,50)(220,50)\LongArrow(330,50)(290,50)\Text(275,50)[]{$-$}
 \Text(240,60)[]{$k-q$}\Text(310,60)[]{$k$}

 \Text(165,-10)[]{$(d)$}

 \end{picture}
 \end{center}

 \caption[]{a) the diagram contributing to $\Phi_\Lambda^{(0)}$; b)
 prescriptions for quark and gluon lines used in Section~4; c)
 graphic representation of the decompositions of $D^{r}_a$; d) the Ward identities in a graphic form.}

 \end{figure}
 calculated  with the following changes with respect to
 the Feynman rules used before (we adopt these changes everywhere in this section)
 \begin{enumerate}

 \item $-igt^a$- factors are removed from all QCD vertices as well as
 $-ieq_fI$ from  QED ones so that nothing except corresponding Dirac
 $\gamma$- matrix remains in any vertex;

 \item $iI$ from quark propagators and $-i\delta^{ab}$ from  gluon ones (they
 do not appear in $D^{r}_a$ but they are present  in other diagrams for $\Phi_\Lambda$)
 are also removed as it is shown in Fig.~5(b);

 \item the factor $-1$ corresponding to the quark loop is omitted;

 \item in quark-Reggeon vertices  Reggeons  are replaced by  gluons with polarization
 vectors equal $p_2^\prime$ (as it is explicitly indicated in the Fig.~5(a)),  since
 the Reggeon interacts with quarks exactly as the gluon with such polarization. Remind
 that for photon polarizations the substitution (\ref{211}) is assumed.

 \end{enumerate}
 Using the relation~(\ref{42}) together with the above agreement for the Feynman
 rules, the amplitude $D^{r}_a$ can be decomposed as it is shown at Fig.~5(c).
 Then we notice that the last term in Fig.~5(c) with  external gluon polarizations
 $q_\perp$ can be omitted since it does not generate a growing with $\tilde s$
 contribution required by Eq.~(\ref{44}). This is because
 to obtain the growing contribution at limited transverse
 polarization vectors of the external gluons one needs purely
 gluonic intermediate states in the $t$- channel. So, $D^{r}_a$ can be presented by the last
 but one term in the graphic relation of Fig.~5(c). To investigate further
 its high energy behaviour we apply the Ward identities shown in a graphic
 form at Fig.~5(d). They allow to present  $D^{r}_a$ in the form given by
 the first equality of Fig.~6(a).

 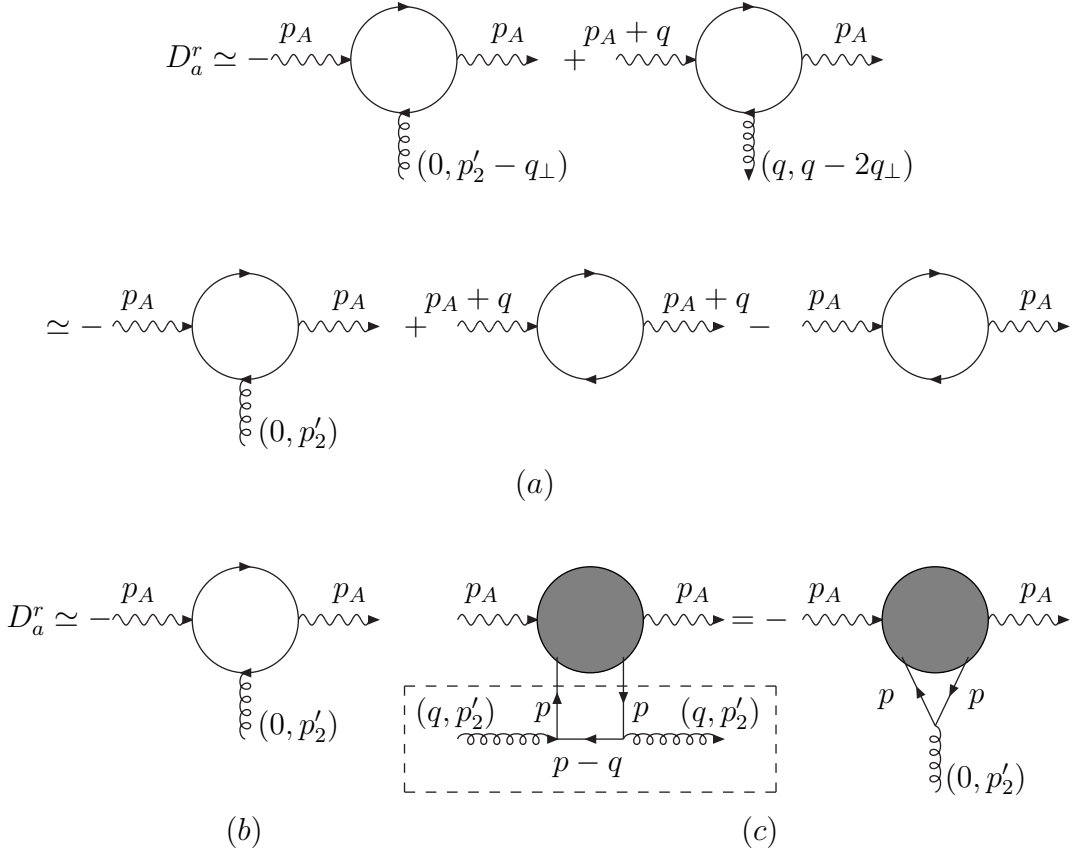
\begin{figure}[tb]
 \vspace{-1.7cm}

 \begin{center}
 \begin{picture}(280,100)(0,0)

 \Photon(40,50)(67,50){2}{4}\ArrowLine(67,50)(70,50)
 \Photon(110,50)(137,50){2}{4}\ArrowLine(137,50)(140,50)
 \BCirc(90,50){20}\ArrowLine(88.5,70)(91.5,70)\ArrowLine(91.5,30)(88.5,30)
 \Gluon(90,5)(90,30){2}{4}\Text(50,60)[]{$p_A$}\Text(130,60)[]{$p_A$}
 \Text(95,10)[l]{$(0, p_2^\prime-q_\perp)$}\Text(0,50)[l]{$D^{r}_a\simeq-$}

 \Photon(170,50)(197,50){2}{4}\ArrowLine(197,50)(200,50)
 \Photon(240,50)(267,50){2}{4}\ArrowLine(267,50)(270,50)
 \BCirc(220,50){20}\ArrowLine(218.5,70)(221.5,70)\ArrowLine(221.5,30)(218.5,30)
 \Gluon(220,30)(220,8){2}{4}\ArrowLine(220,8)(220,5)\Text(175,60)[]{$p_A+q$}
 \Text(260,60)[]{$p_A$}\Text(225,10)[l]{$(q, q-2q_\perp)$}\Text(155,50)[]{$+$}

 \end{picture}
 \begin{picture}(400,100)(0,0)

 \Photon(40,50)(67,50){2}{4}\ArrowLine(67,50)(70,50)
 \Photon(110,50)(137,50){2}{4}\ArrowLine(137,50)(140,50)
 \BCirc(90,50){20}\ArrowLine(88.5,70)(91.5,70)\ArrowLine(91.5,30)(88.5,30)
 \Gluon(90,5)(90,30){2}{4}\Text(50,60)[]{$p_A$}\Text(130,60)[]{$p_A$}
 \Text(95,10)[l]{$(0, p_2^\prime)$}\Text(15,50)[l]{$\simeq-$}

 \Photon(170,50)(197,50){2}{4}\ArrowLine(197,50)(200,50)
 \Photon(240,50)(267,50){2}{4}\ArrowLine(267,50)(270,50)
 \BCirc(220,50){20}\ArrowLine(218.5,70)(221.5,70)\ArrowLine(221.5,30)(218.5,30)
 \Text(175,60)[]{$p_A+q$}\Text(265,60)[]{$p_A+q$}\Text(155,50)[]{$+$}

 \Photon(300,50)(327,50){2}{4}\ArrowLine(327,50)(330,50)
 \Photon(370,50)(397,50){2}{4}\ArrowLine(397,50)(400,50)
 \BCirc(350,50){20}\ArrowLine(348.5,70)(351.5,70)\ArrowLine(351.5,30)(348.5,30)
 \Text(310,60)[]{$p_A$}\Text(390,60)[]{$p_A$}\Text(285,50)[]{$-$}

 \Text(200,-10)[]{$(a)$}

 \end{picture}
 \begin{picture}(400,110)(0,0)

 \Photon(40,50)(67,50){2}{4}\ArrowLine(67,50)(70,50)
 \Photon(110,50)(137,50){2}{4}\ArrowLine(137,50)(140,50)
 \BCirc(90,50){20}\ArrowLine(88.5,70)(91.5,70)\ArrowLine(91.5,30)(88.5,30)
 \Gluon(90,5)(90,30){2}{4}\Text(50,60)[]{$p_A$}\Text(130,60)[]{$p_A$}
 \Text(95,10)[l]{$(0, p_2^\prime)$}\Text(0,50)[l]{$D^{r}_a\simeq-$}
 \Text(90,-30)[]{$(b)$}

 \Photon(170,50)(197,50){2}{4}\ArrowLine(197,50)(200,50)
 \Photon(240,50)(267,50){2}{4}\ArrowLine(267,50)(270,50)
 \GCirc(220,50){20}{0.5}\ArrowLine(207.5,5)(207.5,36)
 \ArrowLine(232.5,36)(232.5,5)\ArrowLine(232.5,5)(207.5,5)
 \Text(180,60)[]{$p_A$}\Text(260,60)[]{$p_A$}
 \ArrowLine(204.5,5)(207.5,5)\ArrowLine(267,5)(270,5)
 \Gluon(170,5)(204.5,5){2}{6}\Gluon(232.5,5)(267.5,5){2}{6}
 \Text(170,15)[]{$(q, p_2^\prime)$}\Text(270,15)[]{$(q, p_2^\prime)$}
 \Text(200,15)[l]{$p$}\Text(243,15)[r]{$p$}\Text(220,-5)[]{$p-q$}
 \DashLine(150,-15)(290,-15){4}\DashLine(150,25)(290,25){4}
 \DashLine(150,-15)(150,25){4}\DashLine(290,-15)(290,25){4}

 \Photon(300,50)(327,50){2}{4}\ArrowLine(327,50)(330,50)
 \Photon(370,50)(397,50){2}{4}\ArrowLine(397,50)(400,50)
 \GCirc(350,50){20}{0.5}\Text(310,60)[]{$p_A$}\Text(390,60)[]{$p_A$}
 \Text(285,50)[]{$=-$}\Gluon(350,10)(350,-15){2}{4}
 \ArrowLine(350,10)(337.5,36)\ArrowLine(362.5,36)(350,10)
 \Text(330,20)[l]{$p$}\Text(370,20)[r]{$p$}
 \Text(355,-10)[l]{$(0, p_2^\prime)$}\Text(285,-30)[]{$(c)$}

 \end{picture}
 \end{center}
 \vspace{1.0cm}

 \caption[]{a) decompositions of  $D^{r}_a$ with the use of the Ward
 identities; b) the final result for the high energy asymptotics of  $D^{r}_a$;
 c) the illustration of this final result.}

 \end{figure}

 We then notice that the parts of the
 diagrams in this equality related to the $q_\perp$- polarizations can be
 omitted because the corresponding contributions do not grow with $\tilde s$.
 This admits to apply again the Ward identities and obtain the second equality at  Fig.~6(a), where
 the last diagram evidently does not grow with $\tilde s$ and can be omitted.

 As for the last but one diagram of Fig.~6(a) we note, first of all, that
 the external virtual photons there have different
 from  $p_A$ momenta (as it is indicated explicitly),
 so that the corresponding amplitude can  grow with $\tilde s$. Nevertheless, it can be omitted also.
 The reason is that the only energy scale for this amplitude is just $\tilde s$ and therefore its
 high energy behaviour in the used by us dimensional regularization is fixed to be
 $\left( -\tilde s \right)^{\epsilon+1}$,  so that its contribution to~(\ref{44})
 is proportional to the following expression
 \begin{equation}\label{45}
 \int_{C_M}\frac{d\tilde s}{2\pi i}\left( -\tilde s \right)^{\epsilon-1}
 = -\int_0^{M^2\rightarrow\infty}\frac{d\tilde s}{2\pi i}\Delta_{\tilde s}\left(
 -\tilde s \right)^{\epsilon-1} = -\frac{\sin(\pi\epsilon)}{\pi}\int_0^\infty
 d\tilde s\left( \tilde s \right)^{\epsilon-1} = 0~.
 \end{equation}
 It vanishes since we first tend $M$ to infinity and only after that
 $\epsilon$ goes to $0$, as it is done  systematically in the BFKL approach
 (see Refs.~\cite{LF89}~-~\cite{FFQ94}, for instance). Of course, these two limits must
 commute in final infrared stable results for observables, but at intermediate steps
 the order of the limits adopted from the beginning must be kept the same everywhere.

 We come therefore to the result for the high energy asymptotics of  $D^{r}_a$
 shown graphically in Fig.~6(b). This result demonstrates the complete
 independence of  $D^{r}_a$  from the Reggeon transverse momentum $q_\perp$.
 This means the independence from  $q_\perp$ also for  $\Phi_\Lambda^{(0)}$
 itself, and, as we will shortly see, this property is valid for the complete
 NLO $\Phi_\Lambda$. Here we make one more remark which will be used in the
 following: the diagram for $D^{r}_a$ has a form of the diagram of Fig.~6(c)
 where the marked by a dash box piece can be written down as follows
 \begin{equation}\label{46}
 \frac{\not p_2^\prime\left( \not p - \not q \right)\not p_2^\prime}{(p-q)^2 +
 i\delta} = - \not p_2^\prime + \not p_2^\prime\frac{\left( \vec p - \vec q \right)
 ^2 - \alpha\beta s_1}{(1-\alpha)\beta s_1 + \left( \vec p - \vec q \right)^2 -
 i\delta}~,
 \end{equation}
 in the Sudakov variables
 \begin{equation}\label{47}
 p = \beta p_1 + \alpha p_2^\prime + p_\perp~.
 \end{equation}
 At limited transverse  momentum
 \begin{equation}\label{48}
 \vec p^{~2} \sim \vec q^{~2} \sim Q^2~,
 \end{equation}
 the essential integration region in the longitudinal subspace is
 \begin{equation}\label{49}
 \alpha \sim \vec q^{~2}/s_1 \ll 1~,\ \ \ \beta \sim 1~.
 \end{equation}
 In this region the last term of  Eq.~(\ref{46}) is suppressed and can be
 omitted so that the expression inside the dash box of Fig.~6(c) becomes
 just
 \begin{equation}\label{410}
 \frac{\not p_2^\prime\left( \not p - \not q \right)\not p_2^\prime}{(p-q)^2 +
 i\delta} \rightarrow - \not p_2^\prime~.
 \end{equation}
 As for the region of integration over large transverse momentum, it always brings the
 energy dependence like that of the Eq.~(\ref{45}) and can be also neglected.
 The prescription~(\ref{410}) provides the relation depicted schematically
 in Fig.~6(c) (of course, it reproduces the correct result for $D^{r}_a$).

 We now turn to  consideration of the complete NLO $\Phi_\Lambda$ which can be
 written as the sum of three contributions
 \begin{equation}\label{411}
 \Phi_\Lambda = \Phi_\Lambda^{(0)} + \Phi_\Lambda^{(1)(a)}
 + \Phi_\Lambda^{(1)(na)}~,
 \end{equation}
 where the LO part $\Phi_\Lambda^{(0)}$ is given by~(\ref{44}) and the
 NLO parts are expressed in terms of $9$ diagrams of Fig.~7

 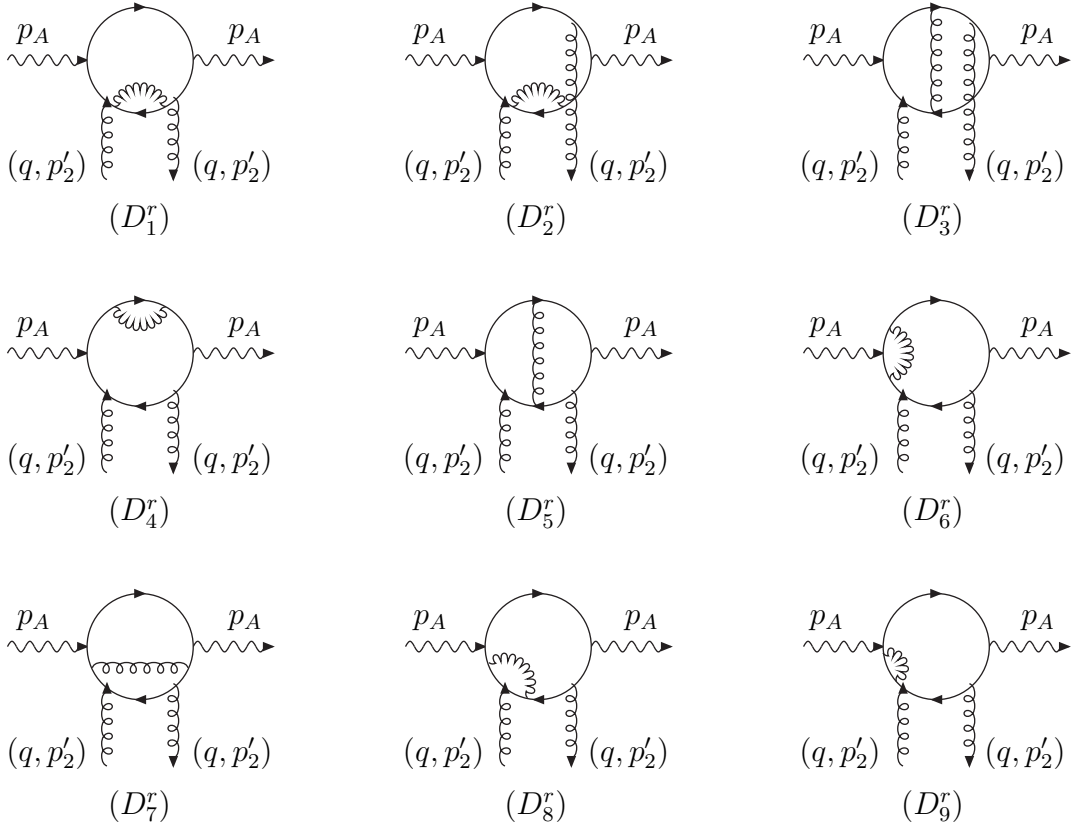
\begin{figure}[tb]
 \vspace{-1.7cm}

 \begin{center}
 \begin{picture}(400,110)(0,0)

 \Photon(0,50)(27,50){2}{4}\ArrowLine(27,50)(30,50)
 \Photon(70,50)(97,50){2}{4}\ArrowLine(97,50)(100,50)
 \BCirc(50,50){20}\ArrowLine(48.5,70)(51.5,70)\ArrowLine(51.5,30)(48.5,30)
 \Gluon(37.5,5)(37.5,33){2}{4}\ArrowLine(37.5,33)(37.5,36)
 \Gluon(62.5,36)(62.5,8){2}{4}\ArrowLine(62.5,8)(62.5,5)
 \Text(10,60)[]{$p_A$}\Text(90,60)[]{$p_A$}\Text(15,10)[]{$(q, p_2^\prime)$}
 \Text(85,10)[]{$(q, p_2^\prime)$}\Text(50,-10)[]{$(D^{r}_1)$}
 \GlueArc(50,30)(10,165,15){-1.5}{8}

 \Photon(150,50)(177,50){2}{4}\ArrowLine(177,50)(180,50)
 \Photon(220,50)(247,50){2}{4}\ArrowLine(247,50)(250,50)
 \BCirc(200,50){20}\ArrowLine(198.5,70)(201.5,70)\ArrowLine(201.5,30)(198.5,30)
 \Gluon(187.5,5)(187.5,33){2}{4}\ArrowLine(187.5,33)(187.5,36)
 \Gluon(212.5,64)(212.5,8){2}{9}\ArrowLine(212.5,8)(212.5,5)
 \Text(160,60)[]{$p_A$}\Text(240,60)[]{$p_A$}\Text(200,-10)[]{$(D^{r}_2)$}
 \Text(165,10)[]{$(q, p_2^\prime)$}\Text(235,10)[]{$(q, p_2^\prime)$}
 \GlueArc(200,30)(10,165,15){-1.5}{8}

 \Photon(300,50)(327,50){2}{4}\ArrowLine(327,50)(330,50)
 \Photon(370,50)(397,50){2}{4}\ArrowLine(397,50)(400,50)
 \BCirc(350,50){20}\ArrowLine(348.5,70)(351.5,70)\ArrowLine(351.5,30)(348.5,30)
 \Gluon(337.5,5)(337.5,33){2}{4}\ArrowLine(337.5,33)(337.5,36)
 \Gluon(362.5,64)(362.5,8){2}{9}\ArrowLine(362.5,8)(362.5,5)
 \Text(310,60)[]{$p_A$}\Text(390,60)[]{$p_A$}\Text(350,-10)[]{$(D^{r}_3)$}
 \Text(315,10)[]{$(q, p_2^\prime)$}\Text(385,10)[]{$(q, p_2^\prime)$}
 \Gluon(350,30)(350,70){2}{6}

 \end{picture}
 \begin{picture}(400,110)(0,0)

 \Photon(0,50)(27,50){2}{4}\ArrowLine(27,50)(30,50)
 \Photon(70,50)(97,50){2}{4}\ArrowLine(97,50)(100,50)
 \BCirc(50,50){20}\ArrowLine(48.5,70)(51.5,70)\ArrowLine(51.5,30)(48.5,30)
 \Gluon(37.5,5)(37.5,33){2}{4}\ArrowLine(37.5,33)(37.5,36)
 \Gluon(62.5,36)(62.5,8){2}{4}\ArrowLine(62.5,8)(62.5,5)
 \Text(10,60)[]{$p_A$}\Text(90,60)[]{$p_A$}\Text(15,10)[]{$(q, p_2^\prime)$}
 \Text(85,10)[]{$(q, p_2^\prime)$}\Text(50,-10)[]{$(D^{r}_4)$}
 \GlueArc(50,70)(10,-15,-165){-1.5}{8}

 \Photon(150,50)(177,50){2}{4}\ArrowLine(177,50)(180,50)
 \Photon(220,50)(247,50){2}{4}\ArrowLine(247,50)(250,50)
 \BCirc(200,50){20}\ArrowLine(198.5,70)(201.5,70)\ArrowLine(201.5,30)(198.5,30)
 \Gluon(187.5,5)(187.5,33){2}{4}\ArrowLine(187.5,33)(187.5,36)
 \Gluon(212.5,36)(212.5,8){2}{4}\ArrowLine(212.5,8)(212.5,5)
 \Text(160,60)[]{$p_A$}\Text(240,60)[]{$p_A$}\Text(200,-10)[]{$(D^{r}_5)$}
 \Text(165,10)[]{$(q, p_2^\prime)$}\Text(235,10)[]{$(q, p_2^\prime)$}
 \Gluon(200,30)(200,70){2}{6}

 \Photon(300,50)(327,50){2}{4}\ArrowLine(327,50)(330,50)
 \Photon(370,50)(397,50){2}{4}\ArrowLine(397,50)(400,50)
 \BCirc(350,50){20}\ArrowLine(348.5,70)(351.5,70)\ArrowLine(351.5,30)(348.5,30)
 \Gluon(337.5,5)(337.5,33){2}{4}\ArrowLine(337.5,33)(337.5,36)
 \Gluon(362.5,36)(362.5,8){2}{4}\ArrowLine(362.5,8)(362.5,5)
 \Text(310,60)[]{$p_A$}\Text(390,60)[]{$p_A$}\Text(315,10)[]{$(q, p_2^\prime)$}
 \Text(385,10)[]{$(q, p_2^\prime)$}\Text(350,-10)[]{$(D^{r}_6)$}
 \GlueArc(330,50)(10,75,-75){-1.5}{8}

 \end{picture}
 \begin{picture}(400,110)(0,0)

 \Photon(0,50)(27,50){2}{4}\ArrowLine(27,50)(30,50)
 \Photon(70,50)(97,50){2}{4}\ArrowLine(97,50)(100,50)
 \BCirc(50,50){20}\ArrowLine(48.5,70)(51.5,70)\ArrowLine(51.5,30)(48.5,30)
 \Gluon(37.5,5)(37.5,33){2}{4}\ArrowLine(37.5,33)(37.5,36)
 \Gluon(62.5,36)(62.5,8){2}{4}\ArrowLine(62.5,8)(62.5,5)
 \Text(10,60)[]{$p_A$}\Text(90,60)[]{$p_A$}\Text(15,10)[]{$(q, p_2^\prime)$}
 \Text(85,10)[]{$(q, p_2^\prime)$}\Text(50,-10)[]{$(D^{r}_7)$}
 \Gluon(32,42)(68,42){2}{6}

 \Photon(150,50)(177,50){2}{4}\ArrowLine(177,50)(180,50)
 \Photon(220,50)(247,50){2}{4}\ArrowLine(247,50)(250,50)
 \BCirc(200,50){20}\ArrowLine(198.5,70)(201.5,70)\ArrowLine(201.5,30)(198.5,30)
 \Gluon(187.5,5)(187.5,33){2}{4}\ArrowLine(187.5,33)(187.5,36)
 \Gluon(212.5,36)(212.5,8){2}{4}\ArrowLine(212.5,8)(212.5,5)
 \Text(160,60)[]{$p_A$}\Text(240,60)[]{$p_A$}\Text(200,-10)[]{$(D^{r}_8)$}
 \Text(165,10)[]{$(q, p_2^\prime)$}\Text(235,10)[]{$(q, p_2^\prime)$}
 \GlueArc(187.5,36)(10,127,-32){-1.5}{8}

 \Photon(300,50)(327,50){2}{4}\ArrowLine(327,50)(330,50)
 \Photon(370,50)(397,50){2}{4}\ArrowLine(397,50)(400,50)
 \BCirc(350,50){20}\ArrowLine(348.5,70)(351.5,70)\ArrowLine(351.5,30)(348.5,30)
 \Gluon(337.5,5)(337.5,33){2}{4}\ArrowLine(337.5,33)(337.5,36)
 \Gluon(362.5,36)(362.5,8){2}{4}\ArrowLine(362.5,8)(362.5,5)
 \Text(310,60)[]{$p_A$}\Text(390,60)[]{$p_A$}\Text(315,10)[]{$(q, p_2^\prime)$}
 \Text(385,10)[]{$(q, p_2^\prime)$}\Text(350,-10)[]{$(D^{r}_9)$}
 \GlueArc(332,42)(6,110,-60){-1.5}{5}

 \end{picture}
 \end{center}

 \caption[]{The diagrams contributing to  $\Phi_\Lambda^{(1)}$.}

 \end{figure}

 $$
 \Phi_\Lambda^{(1)(a)} = - \left( eq_fg^2 \right)^2\frac{C_F\sqrt{N_c^2-1}}{4\pi}
 $$
 $$
 \times\int_{C_M}\frac{d\tilde s}{s_1^2}\left[ D^r_1 + 2D^r_2 + D^r_3 + D^r_4
 + D^{r}_5 + 2D^{r}_6 + D^{r}_7 + 2\left( D^{r}_8 + D^{r}_9 \right) \right]~,
 $$
 \begin{equation}\label{412}
 \Phi_\Lambda^{(1)(na)} = \left( eq_fg^2 \right)^2\frac{C_A\sqrt{N_c^2-1}}
 {8\pi }\int_{C_M}\frac{d\tilde s}{s_1^2}\left[ D^{r}_3 + D^{r}_5 + 2D^{r}_8 \right]~,
 \end{equation}
 with the standard notations $C_F$ and $C_A$ for the Casimir operators in the
 fundamental and adjoint $SU(N_c)$ colour group representations respectively. For
 the first three diagrams of  Fig.~7 the region of integration over limited transverse
 momenta does not give a growing with $\tilde s$ contribution so
 that they separately can be omitted due to relations similar to~(\ref{45}):
 \begin{equation}\label{413}
 \int_{C_M}\frac{d\tilde s}{s_1^2}D^{r}_1 =\int_{C_M}\frac{d\tilde s}{s_1^2}
 D^{r}_2 =\int_{C_M}\frac{d\tilde s}{s_1^2} D^{r}_3 = 0~.
 \end{equation}
 For other diagrams of  Fig.~7, using the Ward identities, one can obtain
 the relations presented in graphic form by Fig.~8.

 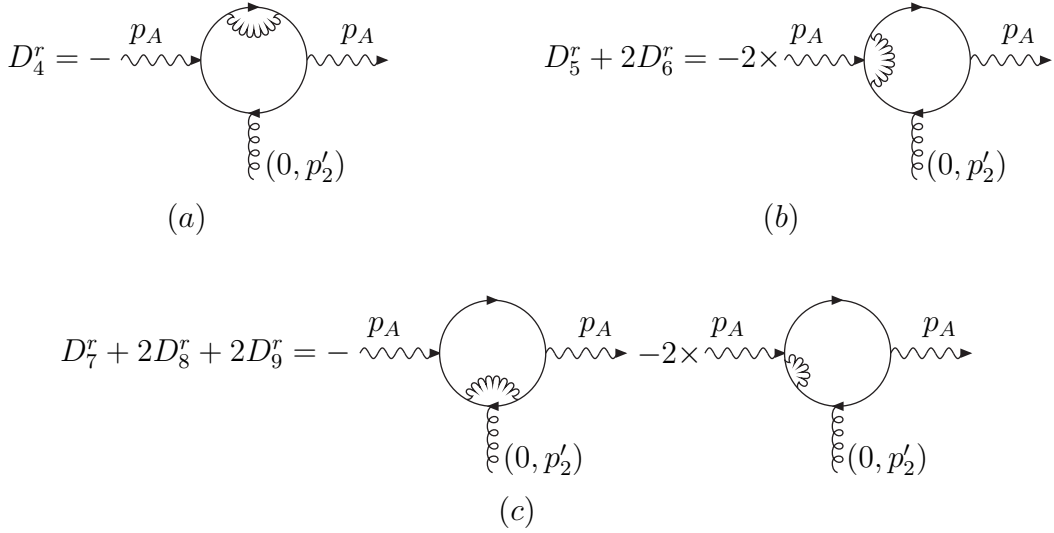
\begin{figure}[tb]
 \vspace{-1.7cm}

 \begin{center}
 \begin{picture}(400,110)(0,0)

 \Photon(50,50)(77,50){2}{4}\ArrowLine(77,50)(80,50)
 \Photon(120,50)(147,50){2}{4}\ArrowLine(147,50)(150,50)
 \BCirc(100,50){20}\ArrowLine(98.5,70)(101.5,70)\ArrowLine(101.5,30)(98.5,30)
 \Gluon(100,5)(100,30){2}{4}\GlueArc(100,70)(10,-15,-165){-1.5}{8}
 \Text(60,60)[]{$p_A$}\Text(140,60)[]{$p_A$}\Text(120,10)[]{$(0, p_2^\prime)$}
 \Text(47,50)[r]{$D^{r}_4=-$}\Text(75,-10)[]{$(a)$}

 \Photon(300,50)(327,50){2}{4}\ArrowLine(327,50)(330,50)
 \Photon(370,50)(397,50){2}{4}\ArrowLine(397,50)(400,50)
 \BCirc(350,50){20}\ArrowLine(348.5,70)(351.5,70)\ArrowLine(351.5,30)(348.5,30)
 \Gluon(350,5)(350,30){2}{4}\GlueArc(330,50)(10,75,-75){-1.5}{8}
 \Text(310,60)[]{$p_A$}\Text(390,60)[]{$p_A$}\Text(370,10)[]{$(0, p_2^\prime)$}
 \Text(300,50)[r]{$D^{r}_5+2D^{r}_6=-2\times$}\Text(300,-10)[]{$(b)$}

 \end{picture}
 \begin{picture}(340,110)(60,0)

 \Photon(170,50)(197,50){2}{4}\ArrowLine(197,50)(200,50)
 \Photon(240,50)(267,50){2}{4}\ArrowLine(267,50)(270,50)
 \BCirc(220,50){20}\ArrowLine(218.5,70)(221.5,70)\ArrowLine(221.5,30)(218.5,30)
 \Gluon(220,5)(220,30){2}{4}\GlueArc(220,30)(10,165,15){-1.5}{8}
 \Text(180,60)[]{$p_A$}\Text(260,60)[]{$p_A$}\Text(240,10)[]{$(0, p_2^\prime)$}
 \Text(167,50)[r]{$D^{r}_7+2D^{r}_8+2D^{r}_9=-$}

 \Photon(300,50)(327,50){2}{4}\ArrowLine(327,50)(330,50)
 \Photon(370,50)(397,50){2}{4}\ArrowLine(397,50)(400,50)
 \BCirc(350,50){20}\ArrowLine(348.5,70)(351.5,70)\ArrowLine(351.5,30)(348.5,30)
 \Gluon(350,5)(350,30){2}{4}\GlueArc(332,42)(6,110,-60){-1.5}{5}
 \Text(310,60)[]{$p_A$}\Text(390,60)[]{$p_A$}\Text(370,10)[]{$(0, p_2^\prime)$}
 \Text(300,50)[r]{$-2\times$}\Text(230,-10)[]{$(c)$}

 \end{picture}
 \end{center}

 \caption[]{Graphic relations for the diagrams of the Fig.~7.}

 \end{figure}

  Let us note that the
 diagrams for $D^{r}_4, D^{r}_6, D^{r}_7$ and $D^{r}_9$ are of the kind of Fig.~6(c) and
 satisfy therefore the same relation in their high energy limit. It means that
 $2D^{r}_6$ and $D^{r}_7+2D^{r}_9$ saturate the equalities of
 Figs.~8(b) and (c) correspondingly, so that $D^{r}_5$ and
 $D^{r}_8$ disappear (separately), although it is not so
 evident as in the case of the first three diagrams of Fig.~7.

 The final conclusion of this Section is that the NLO $\Phi_\Lambda$ consists
 of the integrals over the large  circle in the complex $\tilde s$- plane from
 the amplitudes corresponding to the diagrams presented by Figs.~5(a) and 7,
 which either have the structure of the Fig.~6(c) and do not therefore depend
 on $q_\perp$ in the high energy limit or separately disappear in this limit.
 It leads to the independence from the Reggeon transverse momentum also for
 the $\Phi_\Lambda$ itself. This  conclusion is valid in the Feynman gauge
 for virtual gluons which we use here.

 \section{Conclusions}
 \setcounter{equation}{0}

 The virtual photon impact factor $\Phi_{\gamma^*}(\vec q, s_0)$,
 being the impact factor of the colourless object, has an important property
 \begin{equation}\label{52}
 \Phi_{\gamma^*}(\vec q= 0, s_0) = 0~,
 \end{equation}
 that is related to the gauge invariance and is
 necessary for the infrared finiteness of the cross section describing the
 collision of colourless particles. Since the last two terms in (\ref{29}) vanish at $\vec q =0$,
 the contribution $\Phi_M(\vec q)$ has also to possess this property
 \begin{equation}\label{52a}
 \Phi_{M}(\vec q= 0) = \Phi_\Delta
 (\vec q=0) + \Phi_\Lambda(\vec q = 0)=0~.
 \end{equation}
 We  have shown in the previous Section that $\Phi_\Lambda$ does not depend on $\vec q$.
 It means that  $\Phi_\Lambda(q)=-\Phi_\Delta(q=0)$ and $\Phi_M(\vec q)$
 can be presented as  
 \begin{equation}\label{51}
 \Phi_M(\vec q)= \Phi_\Delta
 (\vec q) - \Phi_\Delta(\vec q = 0)~,
 \end{equation}
 where $\Phi_\Delta$ defined by the Eq.~(\ref{313}) and Fig.~4.

 We came to the above simplified representation using the analytical properties
 of the Feynman diagrams of the effective quantum field theory with Reggeized
 gluon included. A part of the diagrams in the expression~(\ref{210}) for the
 impact factor with known high energy behaviour  was moved from the
 integral over the $\tilde s$- channel discontinuity to the integral over the
 infinite circle in the complex squared energy plane. According to our results
 this part serves just as a counterterm restoring the correct small $\vec q$
 behaviour of the other part of the impact factor $\Phi_\Delta$,
 so that it does not have to be calculated explicitly.
 In other words, we have performed
 the cancellation of some irrelevant terms before doing  the real
 calculations.

 Let us note that the approach using analytical
 properties of amplitudes of the effective  field theory  can be useful
 also for  calculation of  other NLO impact factors. It is  analogous to the approach
 which was used in QED for the derivation of the sum rules relating the cross
 sections of various production processes in the Weizsaker-Williams approximation and the
 slope of the Dirac form factor at zero momentum transfer~\cite{KLM}.

 To complete our conclusions let us consider how the approach works in the simplest
 case of the Born virtual photon impact factor. Here we have
 \begin{equation}\label{54}
 \Phi_{\gamma^*}^{(0)}(\vec q) = \Phi_\Delta^{(0)}(\vec q) - \Phi_\Delta^{(0)}(\vec q = 0)~,
 \ \ \ \Phi_\Delta^{(0)}(\vec q) = - i\int_0^\infty\frac{d\tilde s}{2\pi}\Delta_{\tilde s}
 D_{4(a)}~,
 \end{equation}
 where only the first diagram in Fig.~4 contributes. The calculation becomes
 very simple and we just quote the result without presentation of any details
 $$
 \Phi_\Delta^{(0)}(\vec q) = e^2g^2\sqrt{N_c^2-1}\sum_fq_f^2\frac{\Gamma(1-
 \epsilon)}{(4\pi)^{2+\epsilon}}\biggl\{ \left( Q^2 \right)^\epsilon\frac{2}
 {\epsilon}\int_0^1dx\left( x(1-x) \right)^\epsilon
 $$
 $$
 \times\biggl[ (1+\epsilon)\left( 1+\epsilon-2x(1-x) \right)T^{(+)} - 4\epsilon
 x(1-x)L^{(+)} \biggr]
 $$
 $$
 + \int_0^1\int_0^1\frac{dxdy\, \vec q^{~2}}{\left( y(1-y)\vec q^{~2} + x(1-x)Q^2
 \right)^{1-\epsilon}}\biggl[ \left( 1+\epsilon-2x(1-x) \right)
 $$
 $$
 \times\left( 1+\epsilon-2(1+2\epsilon)y(1-y) \right)T^{(+)} + 4x(1-x)\left(
 2(1+2\epsilon)y(1-y)-\epsilon \right)L^{(+)}
 $$
 \begin{equation}\label{55}
 + 4(1+2\epsilon)x(1-x)y(1-y)T^{(-)} \biggr] \biggr\}~.
 \end{equation}
 According  to our approach, Eq.~(\ref{54}), the LO virtual photon impact factor itself is given
 by the above expression without the first term in the curly brackets
 $$
 \Phi_{\gamma^*}^{(0)}(\vec q) = e^2g^2\sqrt{N_c^2-1}\sum_fq_f^2\frac{\Gamma(1-\epsilon)}
 {(4\pi)^{2+\epsilon}}\int_0^1\int_0^1\frac{dxdy\vec q^{~2}}{\left( y(1-y)\vec q
 ^{~2} + x(1-x)Q^2 \right)^{1-\epsilon}}\biggl[ \left( 1+\epsilon \right.
 $$
 $$
 \left.-2x(1-x) \right)\left( 1+\epsilon-2(1+2\epsilon)y(1-y) \right)T^{(+)} + 4x(1-x)\left(
 2(1+2\epsilon)y(1-y)-\epsilon \right)L^{(+)}
 $$
 $$
 + 4(1+2\epsilon)x(1-x)y(1-y)T^{(-)} \biggr] \approx\sqrt{N_c^2-1}  \sum_fq_f^2
 \frac{e^2g^2}{(4\pi)^2}\int_0^1\int_0^1\frac{dxdy\vec q^{~2}}{y(1-y)\vec q^{~2} +
 x(1-x)Q^2}
 $$
 \begin{equation}\label{56}
 \times\biggl[ \left( 1-2x(1-x) \right)\left( 1-2y(1-y) \right)T^{(+)} + 8x(1-x)
 y(1-y)\left( L^{(+)} + \frac{1}{2}T^{(-)} \right) \biggr]~,
 \end{equation}
 where the last approximate equality shows the result in the physical limit
 $\epsilon = 0$ while the first exact in $\epsilon$ equation is necessary in the
 NLO to evaluate the counterterm in ~(\ref{29}).
 It can be checked by direct calculation that the first term in r.h.s of
 Eq.~(\ref{55}) which we subtracted according to our procedure (\ref{54})
 coincides, of course, with the contribution (with the reverse sign) of the diagram
 Fig.~5(a). So, considering $\Phi_{\gamma^*}^{(0)}(\vec q)$
 we need to calculate in a frame of our method only one diagram  instead of
 two.

 It should be mentioned that, in a difference to approach used in
 \cite{FIK1}~-~\cite{22a},
 the method described above implies an integration at the intermediate
 stages over the phase space of an individual cut diagram.
 This integration is divergent in ultraviolet and is regularized in a
 dimensional regularization method. See, for instance, the first term of
 Eq.~(\ref{55}), where the singularity at $\varepsilon\to 0$ is an
 ultraviolet pole. These additional ultraviolet singularities disappear of
 course in the difference of two terms in Eq.~(\ref{51}). At LO this
 cancellation is very simple and results just in removing the first term
 in Eq.~(\ref{55}) which does not depend on $\vec q$. At NLO the situation
 is more complicated since cut diagrams involve one-loop amplitudes that,
 in practice, one needs to expand in $\epsilon$ in order to proceed with subsequent
 calculations.
 At this stage one has to
 use with care the results for cut diagrams reported previously in
 \cite{FIK1}~-~\cite{22a}. In some kinematical regions the procedure
 of the $\varepsilon$- expansion for the one-loop amplitudes may need a revision.

 Eq.~(\ref{56}) restores the
 well known answer~\cite{Cheng} .
 The calculation in the
 NLO, instead, remains to be rather complicated even with the use of the
 simplified expression~(\ref{51}).
 Nevertheless, it is to be done since it
 has a number of the phenomenological applications.  We hope to solve this problem in our
 subsequent publications.

 \vskip 1.5cm \underline {Acknowledgment}: Two of us (M.K. and D.I.) thank the
 Dipartimento di Fisica della Universit\`a della Calabria for the warm
 hospitality while a part of this work was done.
 This work was partly supported by the INTAS (00-0036 and 00-00679)
 and by the Russian Fund of Basic Researches (00-15-96691 and 01-02-16042).
 D.I. was supported by DFG and BMBF (06OR984).
 V.S.F. thanks the Alexander von Humboldt foundation for the
 research award, the Universit\"at Hamburg and DESY for their warm
 hospitality while a part of this work was done.

\end{document}